\documentclass[sigconf]{acmart}

\usepackage{graphicx, subcaption}
\usepackage[justification=centering]{caption}
\usepackage{comment}

\usepackage{balance}

\AtBeginDocument{%
  \providecommand\BibTeX{{%
    \normalfont B\kern-0.5em{\scshape i\kern-0.25em b}\kern-0.8em\TeX}}}

\copyrightyear{2024}
\acmYear{2024}
\setcopyright{rightsretained}
\acmConference[WiSec '24]{Proceedings of the 17th ACM Conference on Security
and Privacy in Wireless and Mobile Networks}{May 27--30, 2024}{Seoul, Republic
of Korea}
\acmBooktitle{Proceedings of the 17th ACM Conference on Security and Privacy
in Wireless and Mobile Networks (WiSec '24), May 27--30, 2024, Seoul, Republic
of Korea}\acmDOI{10.1145/3643833.3656123}
\acmISBN{979-8-4007-0582-3/24/05}

\makeatletter
\gdef\@copyrightpermission{
  \begin{minipage}{0.3\columnwidth}
   \href{https://creativecommons.org/licenses/by/4.0/}{\includegraphics[width=0.90\textwidth]{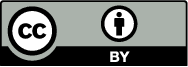}}
  \end{minipage}\hfill
  \begin{minipage}{0.7\columnwidth}
   \href{https://creativecommons.org/licenses/by/4.0/}{This work is licensed under a Creative Commons Attribution International 4.0 License.}
  \end{minipage}
  \vspace{5pt}
}
\makeatother

\begin{document}

%%
%% The "title" command has an optional parameter,
%% allowing the author to define a "short title" to be used in page headers.
\title{Securing Contrastive mmWave-based Human Activity Recognition against Adversarial Label Flipping}

\author{Amit Singha}
\affiliation{%
  \institution{Purdue University}
  \city{West Lafayette}
  \state{IN}
  \country{USA}
}
\email{singha3@purdue.edu}

\author{Ziqian Bi}
\affiliation{%
  \institution{Purdue University}
  \city{West Lafayette}
  \state{IN}
  \country{USA}
}
\email{bi32@purdue.edu}

\author{Tao Li}
\affiliation{%
  \institution{Purdue University}
  \city{West Lafayette}
  \state{IN}
  \country{USA}
}
\email{li4270@purdue.edu}

\author{Yimin Chen}
\affiliation{%
  \institution{University of Massachusetts}
  \city{Lowell}
  \state{MA}
  \country{USA}
}
\email{ian_chen@uml.edu}

\author{Yanchao Zhang}
\affiliation{%
  \institution{Arizona State University}
  \city{Tempe}
  \state{AZ}
  \country{USA}
}
\email{yczhang@asu.edu}

%%
%% By default, the full list of authors will be used in the page
%% headers. Often, this list is too long, and will overlap
%% other information printed in the page headers. This command allows
%% the author to define a more concise list
%% of authors' names for this purpose.
% \renewcommand{\shortauthors}{Trovato and Tobin, et al.}
%\renewcommand{\shortauthors}{Singha et al.}
%\renewcommand{\shortauthors}{Amit Singha, Ziqian Bi, Tao Li, Yimin Chen, \& Yanchao Zhang}

\renewcommand{\shortauthors}{Amit Singha, Ziqian Bi, Tao Li, Yimin Chen, \& Yanchao Zhang}

%%
%% The abstract is a short summary of the work to be presented in the
%% article.
\begin{abstract}

Wireless Human Activity Recognition (HAR), leveraging their non-intrusive nature, has the potential to revolutionize various sectors, including healthcare, virtual reality, and surveillance. The advent of millimeter wave (mmWave) technology has significantly enhanced the capabilities of wireless HAR systems. This paper presents the first systematic study on the vulnerabilities of mmWave-based HAR to label flipping poisoning attacks in the context of supervised contrastive learning. We identify three label poisoning attacks on the contrastive mmWave-based HAR and propose corresponding countermeasures. The efficacy of the attacks and also our countermeasures are experimentally validated on a
prototype system. The attacks and countermeasures can be easily
extended to other wireless HAR systems, thereby promoting security considerations in system design and deployment.

\end{abstract}

%%
%% The code below is generated by the tool at http://dl.acm.org/ccs.cfm.
%% Please copy and paste the code instead of the example below.
%%

%%
%% Keywords. The author(s) should pick words that accurately describe
%% the work being presented. Separate the keywords with commas.

%% A "teaser" image appears between the author and affiliation
%% information and the body of the document, and typically spans the
%% page.
%\begin{teaserfigure}
%  \includegraphics[width=\textwidth]{sampleteaser}
%  \caption{Seattle Mariners at Spring Training, 2010.}
%  \Description{Enjoying the baseball game from the third-base
%  seats. Ichiro Suzuki preparing to bat.}
%  \label{fig:teaser}
%\end{teaserfigure}

%%
%% This command processes the author and affiliation and title
%% information and builds the first part of the formatted document.
\maketitle

\section{Introduction}

Wireless human activity recognition (HAR) has gained significant attention in the past decade as a technology that enables the detection and monitoring of human gesture, behavior, and movement wirelessly \cite{LiuWir19, WangDev17, LiWih20, Li_2022}. It works by detecting and recognizing changes in the wireless signal caused by human activities. Wireless HAR has diverse applications, such as healthcare, virtual reality, monitoring and surveillance, defense and military, and smart buildings. A key benefit of wireless HAR is its non-intrusive nature, along with its ability to function through walls and obstacles, making it ideal for situations where traditional (wearable) sensors or cameras may not work properly.  

\begin{figure}[htbp]
    \centering
    \begin{subfigure}[b]{0.23\textwidth}
        \centering
        \includegraphics[width=\textwidth]{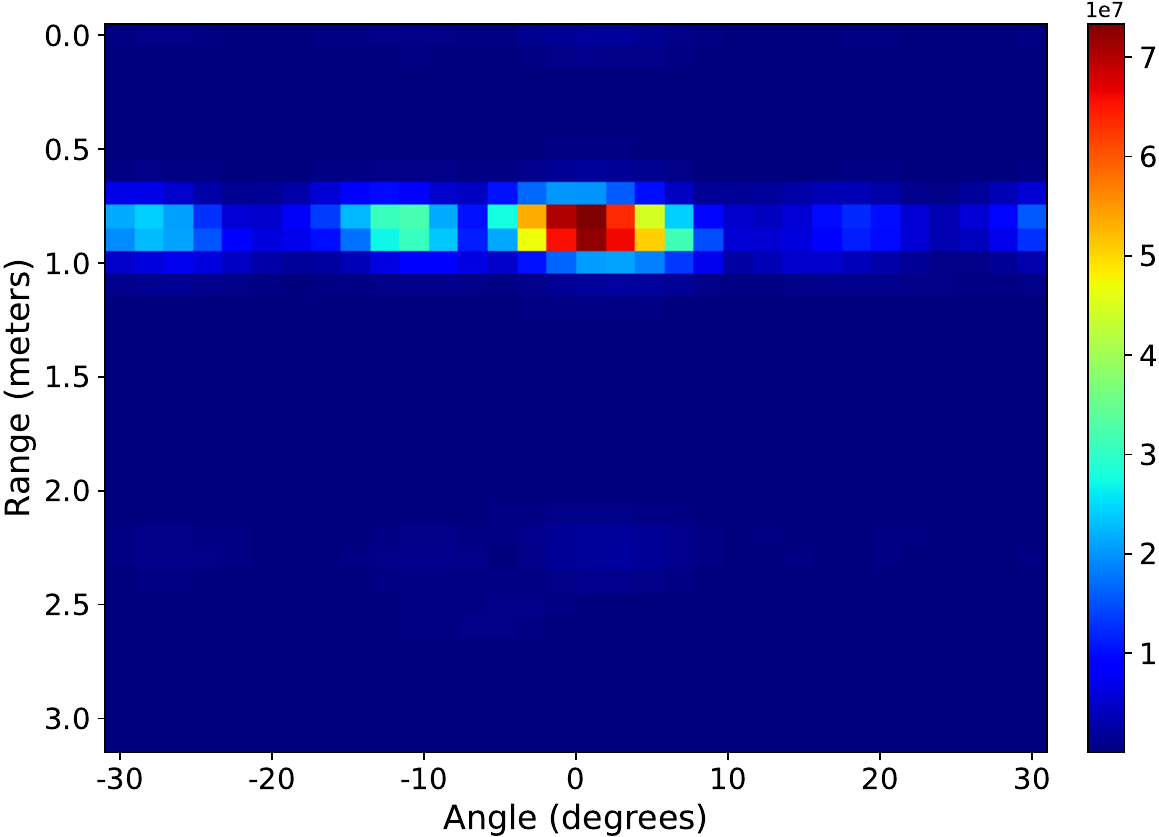}
        \caption{The hand "pull" activity.}
        \label{fig:pull}
    \end{subfigure}
    \hfill
    \begin{subfigure}[b]{0.23\textwidth}
        \centering
        \includegraphics[width=\textwidth]{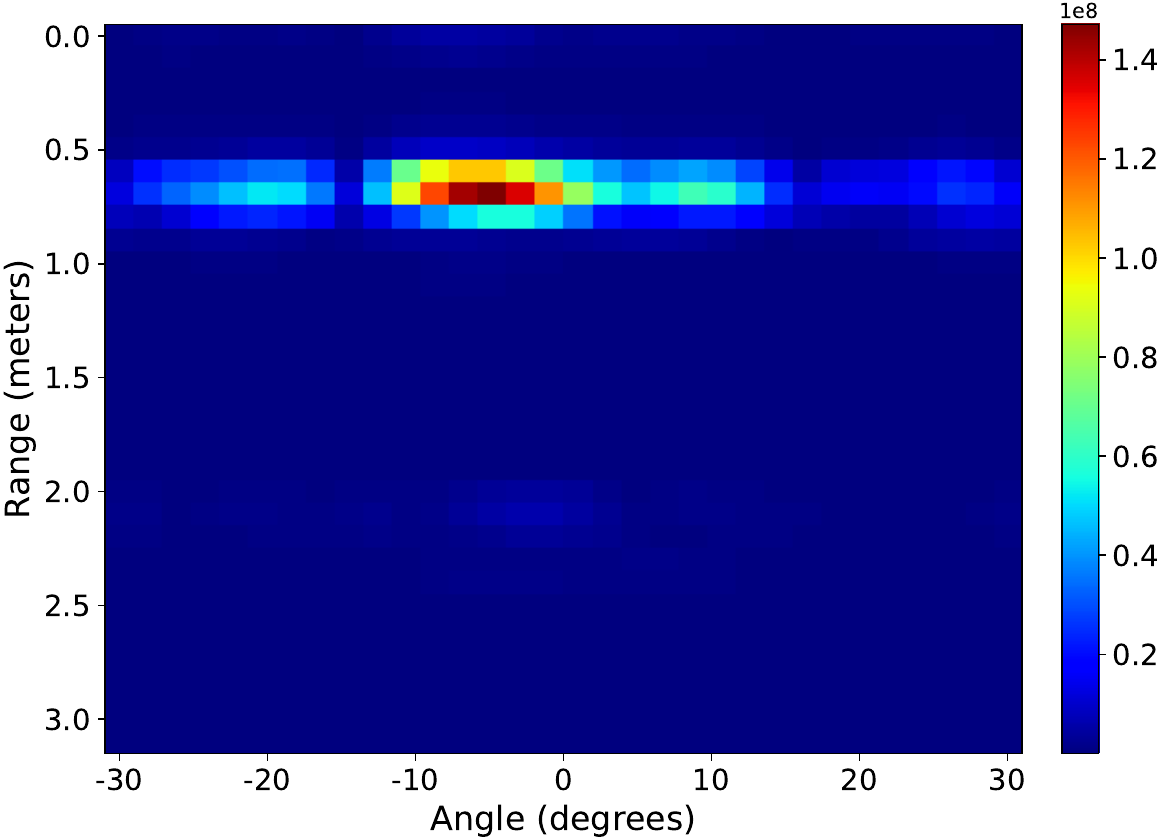}
        \caption{The hand "push" activity.}
        \label{fig:push}
    \end{subfigure}
    \caption{Heatmaps of two activities.}
    \label{fig:twogestures}
\end{figure}

One breakthrough in this field is millimeter wave (mmWave) technology, which operates within the frequency range of 24 GHz to 300 GHz, corresponding to wavelengths between 12.5 mm and 1 mm. The exceptional bandwidth and high-speed capabilities of mmWave have unlocked
new possibilities for HAR applications requiring low latency and high speeds over short distances \cite{ZhangSur23, NiuSur15}. In addition to new wireless techniques, novel AI approaches like contrastive learning have also been used in wireless HAR to improve performance and robustness \cite{HaresCon21,LiWea22,SongRfu22}. Contrastive learning improves data representation by focusing on the similarity between instances, an aspect neglected in conventional machine learning.

Although Wireless HAR, enhanced with cutting-edge wireless and AI technologies, demonstrates remarkable performance, there is still a lack of research on its susceptibility to label flipping poisoning attacks. Label flipping involves altering the labels in the training data to negatively affect the system's overall performance. Most of the existing label flipping attacks target general machine learning models, neglecting to account for the distinctive characteristics in wireless HAR \cite{TianCom22, SongLea22, CinWil23}. For example, detecting label flipping between a "cat" and a "dog" in traditional image data is relatively straightforward. However, identifying such attacks between non-intuitive "pull" and "push" wireless heatmaps, as illustrated in Figure \ref{fig:twogestures}, poses a greater challenge. 

The objective of this work is to identify vulnerabilities in contrastive mmWave-based HAR and propose countermeasures to enhance its \emph{security}, \emph{accuracy}, and \emph{robustness} in the presence of label poisoning attacks. To achieve this goal, we prototype an mmWave-based HAR empowered by supervised contrastive learning to evaluate potential attacks and the corresponding defenses. Although this study focuses on mmWave-based HAR, the rationale can easily extend to other wireless HAR systems, including those based on Wi-Fi and acoustic signals.  We hope that our study can promote security considerations in the early phase of designing and deploying wireless HAR systems.

In contrast to existing research \cite{HanAdv12, ShahiLab22, ShahiAss23}, we propose three novel attacks that exploit similarity of human activity trajectories for label flipping attacks, an area hitherto unexamined. In the first attack, the attacker manipulates the labels randomly to include labels of other activities with both similar and dissimilar trajectories.  In the second attack, the attacker manipulates the label of an activity to match the label of another activity with a different trajectory. In the third attack, the attacker adjusts the label of an activity to correspond with another activity that shares a similar trajectory.

We also develop defense mechanisms to mitigate the impact of malicious labels in HAR systems that rely on mmWave technology and employ supervised contrastive learning (SCL). Our defense mechanism does not require any trusted training dataset assumed by prior research \cite{ShahiLab22, ShahiAss23}.  The basic idea is to identify confident pairs of activities first in the poisoned training dataset using unsupervised contrastive learning. Then, the identified confident pairs are used to find more confident pairs based on representation similarity distribution. Ultimately, we train the mmWave-based HAR model based on all confident activities.

Our contributions can be summarized as follows.

\begin{itemize}
    \item We are the first to study the label flipping poisoning attacks in mmWave-based HAR. Our principles can be easily extended to other wireless HAR systems.

    \item We present three trajectory-based label flipping poisoning attacks and evaluate their performance in a prototype system. We 
    found that activity trajectories do have a large impact on the attacking performance.

    \item We develop defense mechanisms to safeguard mmWave-based HAR systems against adversarial label flipping. Our experimental results demonstrate the robustness of our technique against the identified attacks.
\end{itemize}

The rest of the paper is organized as follows. Section II
introduces a basic mmWave-based HAR system.
Section III describes the prototype system we build based on supervised contrastive learning. Section IV
presents the adversary model and 
three possible attacks. Section V gives our countermeasures
against the identified attacks. Section VI experimentally evaluates
the efficacy of the prototype, attacks, and our countermeasures. Section VII outlines
the related work. Section VIII concludes this paper.

\section{A Basic mmWave-based HAR System}

\begin{figure}[ht]
	\centering
	\includegraphics[width=1\linewidth]{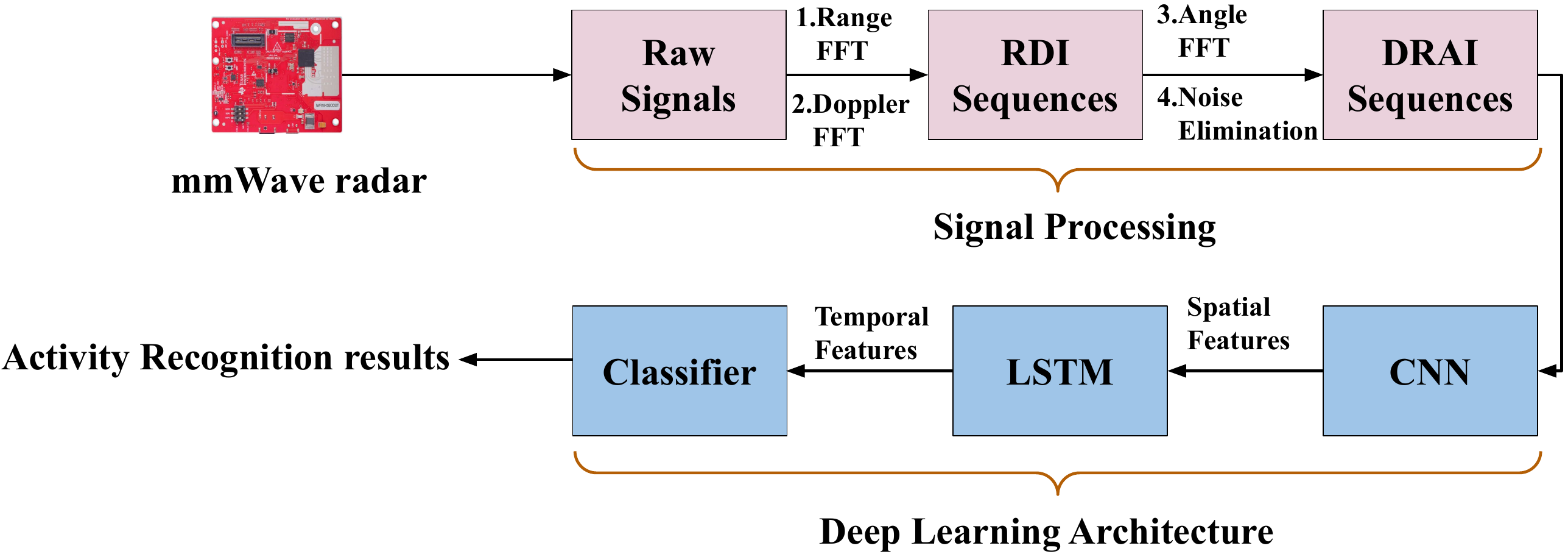}
	\caption{A basic mmWave-based HAR system.}
	\label{fig:Basic_Gesture_Recognition_System}
\end{figure}

In this section, we introduce the basic mmWave-based HAR system that employs supervised learning (SL) with a TI 1843 mmWave radar following the design in \cite{Li_2022, 10001175}. During activity recognition, the transmission antennas of the radar first emit frequency-modulated continuous wave (FMCW)  chirps. The signals are then reflected by each part of the user body and finally received by the receiving antennas of the radar \cite{iovescu2017fundamentals, rao2017introduction}. Then, the transmitted and received signals are mixed to generate the intermediate frequency (IF) signals, which are the input raw signals for the mmWave-based HAR system illustrated in Figure~\ref{fig:Basic_Gesture_Recognition_System}. After that, the system performs Range-FFT and Doppler-FFT to generate the Range Doppler Image (RDI) sequences. RDI sequences are time-series heatmaps that show the range and speed information of objects. The system proceeds to execute Angle-FFT and remove clutters to generate the clean Dynamic Range Angle Image (DRAI) sequences. DRAI sequences are time-series heatmaps that show the range and angle information of objects. 

\begin{figure}[ht]
	\centering
	\includegraphics[width=0.8\linewidth]{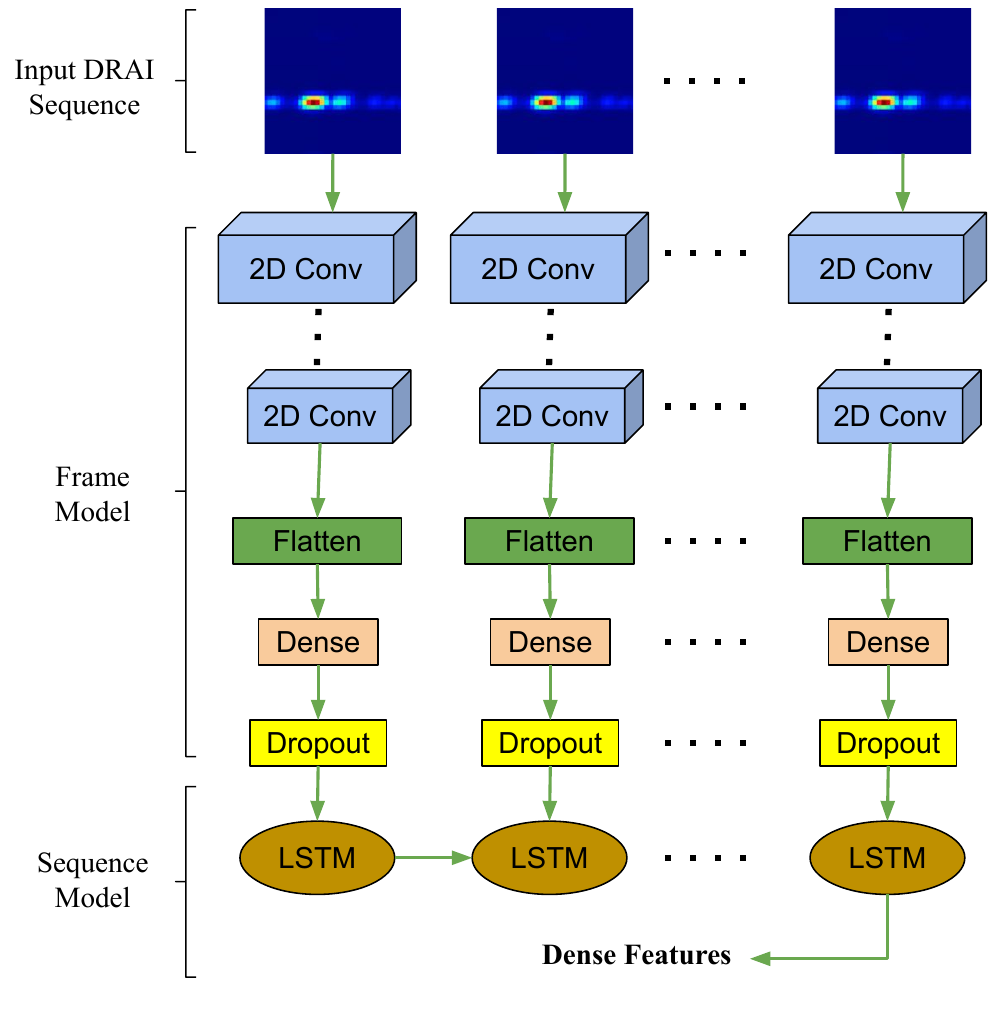}
	\caption{The CNN-LSTM network used for activity recognition.}
	\label{fig:Backbone}
\end{figure}

The system finally uses a hybrid CNN-LSTM model for activity classification, shown in Figure \ref{fig:Backbone}. In particular, CNN  captures spatial features of each heatmap and LSTM extracts temporal features in the time-series heatmaps characterizing the user activity. The CNN architecture comprises three convolutional layers, with channel counts of 16, 32, and 64, respectively. Each layer utilizes a 3x3 kernel size, a stride of 2, and incorporates no padding. The output from the CNN is flattened and then processed through a fully connected layer to yield a condensed feature representation. Subsequently, the LSTM employs a single layer that contains 512 hidden nodes. In the final step, we employ a fully connected layer to classify the feature vector, which encapsulates both spatial and temporal characteristics and produces the ultimate activity recognition result.

Our basic SL-based mmWave HAR system was developed based on a TI 1843 radar, which has three transmission antennas and four receiving antennas. The collected training dataset comprises 10,650 samples, encompassing six distinct hand activities: Push (PH), Pull (PL), Slide Left (LS), Slide Right (RS), Clockwise Turning (CT), and Anticlockwise Turning (AT). These activities were recorded by 25 volunteers, spanning five different locations and six diverse environments, ensuring a rich and varied data pool. The dataset is balanced with 1775 samples of each activity. Our basic system achieves 96.43\% in the final activity recognition.

\section{A Prototype Enhanced by Supervised Contrastive Learning}

%We tackle the noisy labels issue in mmWave-based gesture recognition through a two-part methodology by utilizing Contrastive Learning (CL). In the first part, we develop a baseline by using Supervised Contrastive Learning (SCL)\cite{khosla2021supervised} to improve the gesture classification accuracy compared to Supervised Learning (SL)\cite{Cunningham2008} on the clean labeled mmWave dataset. In the second part, we propose a Selective  Supervised Contrastive Learning (Sel-CL) approach to solve the noisy label issue on mmWave radar data for maintaining high classification accuracy against noisy label data. The purpose of using Sel-CL for handling noisy labeled mmWave data is based on our experiments, where we found that both SL and SCL collapse and do not perform well in the presence of noisy label data. 

%In this first part of our research, we have developed a novel approach for gesture classification on the mmWave-based DRAI dataset by utilizing Supervised Contrastive Learning (SCL) based custom Convolutional Neural Network-Long Short-Term Memory (CNN-LSTM) architecture as depicted in Figure \ref{fig:1st_Project.pdf}. By using this methodology our model can effectively exploit the spatial-temporal characteristics of gesture data.

We now enhance the performance of the basic mmWave-based HAR system through supervised contrastive learning (SCL). Our goal is to train a model capable of producing similar representations for activities within the same class while generating distinct representations for activities across different classes.

\begin{figure}[ht]
	\centering
	\includegraphics[width=\linewidth]{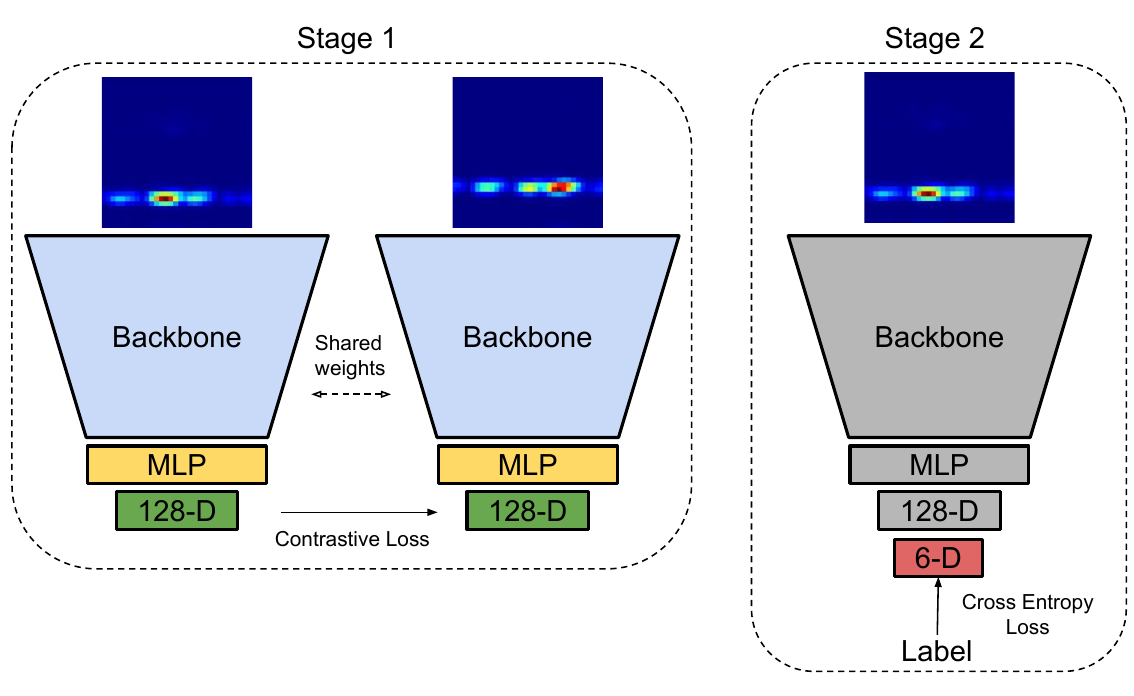}
	\caption{Architecture of supervised contrastive learning (SCL) for enhanced activity recognition.}
	\label{fig:1st_Project.pdf}
\end{figure}

\subsection{Generating Positive and Negative Pairs}
\label{subsec:positive-negative-pairs}

Supervised Contrastive Learning (SCL) leverages label information to enhance the learning of representations in contrastive learning. To train an SCL model, we first need to find positive and negative activity pairs. We randomly select two \emph{different} activities from the same class as a positive pair. The samples in a positive pair teach the model to recognize the underlying similarity between different instances of the same class. In contrast, a negative pair consists of two activity samples from different classes. Samples in a negative pair teach the model to distinguish between dissimilar activities. The goal of SCL model training is to minimize the distance between representations of samples in a positive pair and maximize the distance between representations of samples in a negative pair.

\subsection{Model Training}

Figure \ref{fig:1st_Project.pdf} shows the SCL architecture, which consists of two stages. In the first stage, we design an SCL based CNN-LSTM architecture to capture the spatial and temporal feature vectors of each activity. We refer to the CNN-LSTM architecture as Backbone Network, as illustrated in Figure \ref{fig:Backbone}. Then, we use a Multilayer Perceptron (MLP) to reduce the dimension of the LSTM output into 128-dimensional feature vectors or embedding. For convenience, we denote the embedding of an activity instance \( x_i \) as \( \mathbf{z}_i = f_{\phi}(x_i) \), which serves as the anchor embedding in the contrastive learning framework. The anchor sample \( x_i \) is a specific activity instance from the dataset that acts as a reference for comparing other samples in a batch during the training process. \( f_{\phi} \) symbolizes our model architecture function parameterized by \( \phi \), transforming mmWave hand activity data into meaningful embeddings. Similarly, we use \( \mathbf{z}_p = f_{\phi}(x_p) \) to represent the embedding of a positive sample — a different instance within the same class as the anchor, and \( \mathbf{z}_a = f_{\phi}(x_a) \) for a negative sample from a different activity class.  After that, we apply SCL loss to train our model to learn the representations from an activity. The SCL loss function is designed to ensure that the model learns to produce similar embeddings for activities within the same class (positive pairs) while generating distinct embeddings for activities across different classes (negative pairs), effectively leveraging the contrastive nature of the loss.

The SCL loss function, which is fundamental to our representation learning in the first stage, is formulated as follows:

\begin{equation}
\mathcal{L}_{\text{SCL}} = \sum_{i \in I} -\frac{1}{|P(i)|} \sum_{p \in P(i)} \log \frac{\exp((\mathbf{z}_i \cdot \mathbf{z}_p) / \theta)}{\sum\limits_{a \in A(i)} \exp((\mathbf{z}_i \cdot \mathbf{z}_a) / \theta)},
\end{equation}

\noindent where \( \mathbf{z}_i \), \( \mathbf{z}_p \), \( \mathbf{z}_a \) denote the embedding of anchor, positive, and negative samples, respectively. \( P(i) \) represents the set of all positive samples that share the same label as the anchor. \( A(i) \) includes all samples in the batch, excluding \( i \), which ensures that the anchor is compared against both positive and negative samples. This mechanism effectively handles negative pairs by emphasizing their separation in the embedding space. \( \theta \) is a temperature scaling parameter that controls the separation between positive and negative pairs in the embedding space. The dot product measures the similarity between embeddings. Upon completing the training using the SCL loss, the weights of the entire pipeline, comprising both the backbone and the MLP, are frozen.

%This training focuses on learning embeddings that cluster similar gestures while separating different ones. When the training is completed by using SCL loss the weights of the whole pipeline of backbone with MLP are frozen.

In the second stage of our proposed architecture, we train a linear classifier on top of the frozen Backbone that maps the LSTM output to the predefined activity classes. We use Cross Entropy Loss to train the classifier layer, which is a process of fine-tuning. This completes the entire activity classification process using SCL.

\section{Adversary Model and Attacks}

In this section, we first outline the adversary model and then present three attacks by considering the similarity of activity trajectories.

\subsection{Adversary Model}

The attackers aim to degrade the performance of the wireless HAR system by manipulating the activity labels in the training dataset. This attack differs from other backdoor and adversarial example attacks \cite{GuBad19, YuanAdv19}, as it solely involves altering labels without modifying the data samples themselves. The attackers can be hired by a malicious competitor to ruin the business of the HAR operator, and such instances are not uncommon in the business world. They may also extort the HAR operator or misbehave just for fun.

The training process for a wireless HAR system is labor-intensive and time-consuming. This is primarily due to the necessity of collecting wireless signals from individuals actively engaged in various physical activities. Such a comprehensive collection process inadvertently broadens the attack surface, providing attackers with more opportunities to poison the training dataset. The potential attacking surfaces can be categorized as follows. First, individuals involved in collecting and labeling mmWave signals could intentionally mislabel data. Second, HAR system operators may acquire mmWave training data from personal users or online platforms, which attackers could have mislabeled. Third, the HAR operator may outsource the model training to a third party, which could poison the training data.

\subsection{Attacks}

We observe that some human activities have similar motion trajectories but different motion directions. Based on the similarity of trajectories, we can categorize the six activities in our training data into the following three groups: Push and Pull (back-and-forth motions), Slide Left and Slide Right (lateral movements), and Clockwise Turning and Anticlockwise Turning (circular motions). We aim to study the impact of trajectory similarity on the performance of label flipping attacks in activity classification. In addition, we will also investigate the effect of symmetry during label manipulation. For a symmetric flipping attack, if labels of activity $A$ are altered to represent activity $B$, a corresponding number of labels for activity $B$ are likewise modified to denote activity $A$. In contrast, an asymmetric attack involves altering labels from one activity to another without reciprocation. We study the following three attacks and will evaluate them with symmetric and asymmetric forms in Section 6.

\noindent \textbf{Random attacks:} the attacker randomly chooses a subset of samples within the training data and alters their labels to other arbitrary labels, meaning the flipped labels could represent activities with both similar and dissimilar trajectories.

\noindent \textbf{Across trajectory attacks:} the attacker chooses a few classes of activities within the training data and modifies their labels to labels of other activities with different trajectories.

\noindent \textbf{Inner trajectory attacks:} the attacker also chooses a few classes of activities within the training data and changes their labels to represent activities with similar trajectories. Since the modified label represents an activity with a similar trajectory, the attacker might employ this technique to conceal the malicious label tampering.

\noindent\textbf{Summary of Attacks:} the three attacks above are all in their basic forms. The attacker can conceive many variations of each attack or an arbitrary combination of the three attacks to disrupt the HAR operations. There is thus a pressing need to develop sound defenses to safeguard wireless HAR from these attacks.

\section{Defenses}

We now discuss how to defend against the label flipping attacks by selecting confident activity samples for SCL model training. Below, we represent scalars in lowercase and vectors in bold lowercase. 

\subsection{Overview of Proposed Architecture}
%In our approach of building a malicious label tolerant activity classification model, we leverage a CNN-LSTM architecture, specifically chosen for its ability to capture both the spatial features (through CNN)\cite{o2015introduction} and the temporal dynamics (via LSTM)\cite{yu2019review} inherent in activity data. This combination is particularly effective for analyzing data from mmWave radar, where activitys are characterized not only by their form but also by their evolution over time.

\begin{figure}[ht]
	\centering
	\includegraphics[width=\linewidth]{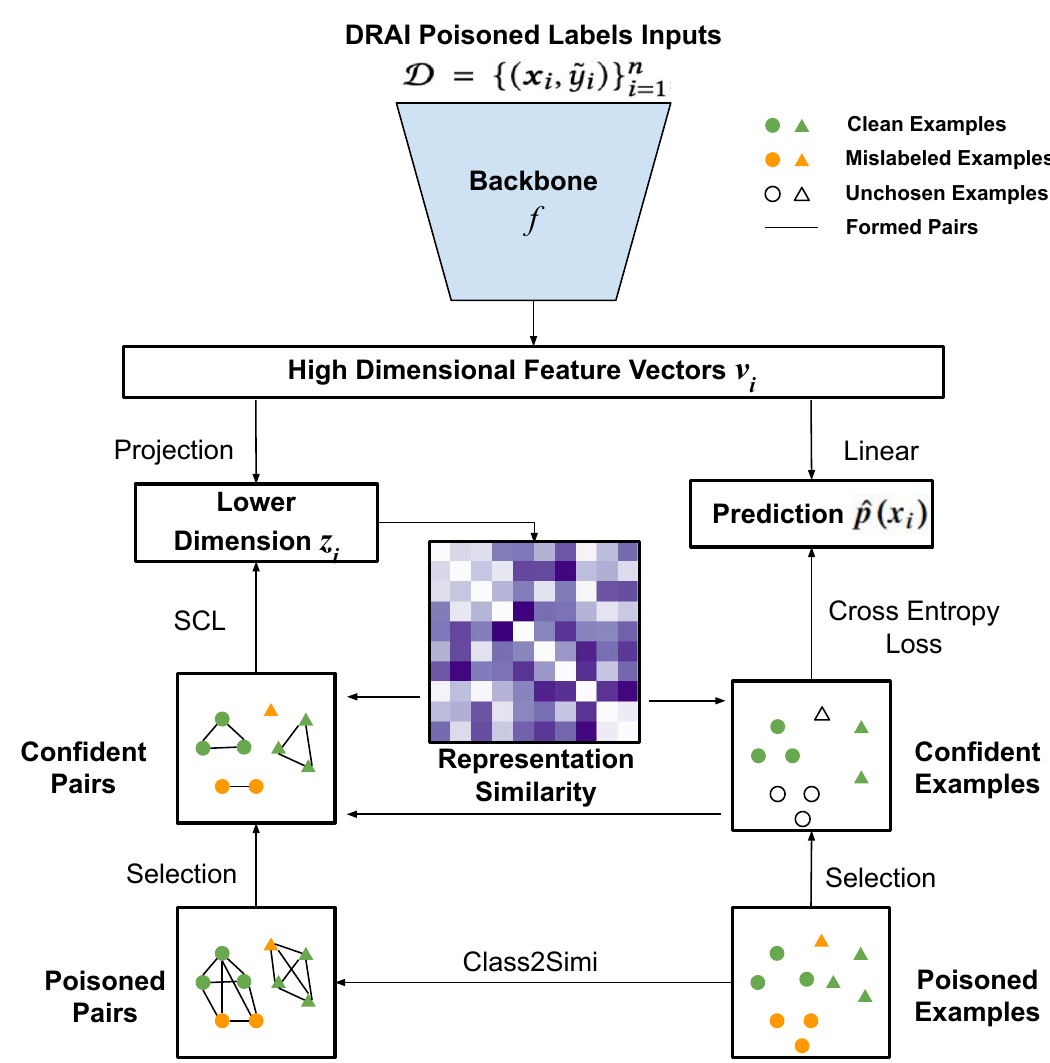}
	\caption{Architecture of the proposed defense against malicious labels for training the SCL model.}
	\label{fig:2nd_Project.pdf}
\end{figure}

Figure \ref{fig:2nd_Project.pdf} illustrates the defense architecture, which \emph{selectively} employs activities in the poisoned training dataset to train the supervised contrastive learning (Sel-CL) model. We adapt the architecture from \cite{li2022selectivesupervised, DBLP:journals/corr/abs-2012-04462} and divide our training process into two phases to address the issue of malicious labels: Pretraining and Fine-Tuning. The main goal of using pretraining in our approach is to learn the more trustworthy representations of activities in the poisoned training data. To achieve this, we extract confident activity labels from the poisoned training dataset and simultaneously select confident pairs from these examples to apply SCL loss on them. We denote the activities in the poisoned training dataset as $\mathcal{D} = \{(\boldsymbol{x}_i, \tilde{y}_i)\}_{i=1}^n$, where $n$ represents the number of samples in the dataset, $\boldsymbol{x}_i$ is the $i$-th activity instance, and $\tilde{y}_i$ represents the untrustworthy label of $\boldsymbol{x}_i$. The true label, albeit sometimes hidden, is represented as $y_i$. 

Our pretraining architecture has three main parts. The first one is a Backbone  \(f\), which transforms each activity instance $\boldsymbol{x}_i$ into a high-dimensional feature vector $\boldsymbol{v}_i$. We employ the identical CNN+LSTM network from our prototype system as the backbone in the defense. The second part is a  classifier head which consists of a fully connected layer. For the prediction of class probabilities \( \boldsymbol{\hat{p}}(\boldsymbol{x}_i) \) we use a softmax activation function. The third part is a projection mechanism which converts  $\boldsymbol{v}_i$ into low dimensional representation \(\boldsymbol{z}_i\). Upon obtaining robust representations of activities through pretraining, we freeze the pretrained backbone during the fine-tuning stage and introduce a new classifier head atop it for the model's activity class predictions.

\subsection{Identifying Confident  Examples and Pairs}

To determine confident pairs, we need to first identify confident examples. For this purpose, we begin training our model using Unsupervised Contrastive Learning (UCL) \cite{chen2020simple} during the initial few epochs. This early training stage is crucial for obtaining low dimensional representations of the activities, which are essential for identifying confident examples in later stages. To determine these confident examples, we measure the agreement between the low dimensional representations and the provided labels by employing cosine distance as a measure of similarity  between two representations \(\boldsymbol{z}_i\) and \(\boldsymbol{z}_j\) as:

\begin{equation}
d(\boldsymbol{z}_i, \boldsymbol{z}_j) = \frac{\boldsymbol{z}_i \cdot \boldsymbol{z}_j}{\|\boldsymbol{z}_i\|\|\boldsymbol{z}_j\|}.
\end{equation}

To correct mislabeled instances in the dataset, we introduce pseudo labels, $\hat{y}_i$, by aggregating the labels from the top-$K$ nearest neighbors of each instance. We select $K = 250$, as the method's efficacy remains stable once $K$ reaches a sufficiently high value. Therefore, we consider the 250 closest counterparts for each data point to determine its pseudo label. This process involves counting the occurrence of each class label among the top-$K$ neighbors and assigning the most frequent class label as the pseudo label $\hat{y}_i$. This method leverages representation similarity to enhance the correction of mislabeled instances, thereby refining the dataset quality.  Through this framework, we approximate the posterior probabilities that an instance $\boldsymbol{x}_i$ belongs to a clean class. We utilize $I[A]$ as an indicator of an event $A$ and $[z]$ to denote the set $\{1, \ldots, z\}$. The equation to approximate these probabilities is given as:

\begin{equation}
\hat{q}_c(\boldsymbol{x}_i) = \frac{1}{K} \sum_{k=1}^{K} I[\hat{y}_k = c] \text{ for } \boldsymbol{x}_k \in N_i \text{ and } c \in [6],
\end{equation} 

\noindent where \(\mathcal{N}_i\) is the collection of \(K\)-nearest instances to \(\boldsymbol{x}_i\). Since our study involves 6 classes, these classes are represented as \(c \in [6]\). Cross-entropy loss is utilized to find confident examples for each class. We designate the collection of confident examples for each class \(c\) as \(\mathcal{T}_c\), which can be formulated as:

\begin{equation}
\mathcal{T}_c = \{(\boldsymbol{x}_i, \tilde{y}_i) \,|\, \text{loss}(\boldsymbol{\hat{q}}(\boldsymbol{x}_i), \tilde{y}_i) < \gamma_c, \, i \in [n] \}, \, c \in [6],
\end{equation} 

\noindent where \(\gamma_c\) represents a specific threshold for each of the 6 classes in our dataset, dynamically set to assure a balanced composition of confident examples across all classes. The appropriate number of examples to select per class is determined by the \(\alpha\)-percentile of agreements between the modified label \(\hat{y}_i\) and the original label \(\tilde{y}_i\) over all classes, calculated as: \( \sum_{i=1}^{n} I[\hat{y}_i = \tilde{y}_i] I[\tilde{y}_i = c], \, c \in [6]. \) Ultimately, we compile the confident example set encompassing all 6 classes, represented as \(\mathcal{T} = \bigcup_{c=1}^{6} \mathcal{T}_c\). In comparison to the original dataset, this set has less noise, making it a more confident base for our training process. 

In the next step, we aim to select confident pairs from the dataset. We define a confident pair as two instances that are similar in their representations and consistent in their labels, as identified in the confident examples set $\mathcal{T}$. Initially, we create a subset of pairs where both instances in each pair have matching labels within $\mathcal{T}$. Subsequently, we expand our selection to include additional pairs based on a dynamically determined threshold \( \gamma \), which helps us identify instances with high representation similarity and consistent pseudo labels. \(\gamma\) is determined by the \(\beta\)-fractile, which is the statistical threshold below which a certain percentage (\(\beta\)\%) of representation similarity scores fall. This ensures the selection of pairs with similarities exceeding \(\gamma\), identifying them as confident pairs.

\subsection{Utilizing Confident Pairs in Representation Learning}

We leverage the confident pairs identified earlier for representation learning and implement supervised contrastive learning \cite{khosla2021supervised}  in each epoch. This method enhances activity representation by concentrating on confident pairs,  strengthening the defense against malicious labels.

Our custom pair selection strategy, detailed in Subsection~\ref{subsec:positive-negative-pairs}, is employed to generate positive and negative pairs for supervised contrastive learning. This approach operates within each mini batch, circumventing the need for data augmentation. Each training mini batch is represented as \(\{(\boldsymbol{x}_i, \tilde{y}_i)\}_{i=1}^{2N}\), where \(i \in I = [1, 2N]\) is the index of a sample in the batch. This procedure of creating positive and negative pairs is crucial for learning a representation space wherein samples of the same class are embedded closer together while those from different classes are positioned further apart.

Given this setup, we apply supervised contrastive learning using the constructed pairs. The contrastive loss \(\mathcal{L}\) is computed as:

\begin{equation}
\mathcal{L} = \sum_{i \in I} -\frac{1}{|\mathcal{G}(i)|} \sum_{g \in \mathcal{G}(i)} \log \left( \frac{\exp(\boldsymbol{z}_i \cdot \boldsymbol{z}_g / \theta)}{\sum_{a \in A(i)} \exp(\boldsymbol{z}_i \cdot \boldsymbol{z}_a / \theta)} \right),
\end{equation}

\noindent where \(A(i)\) is the collection of indices that don't include \(i\), \(\mathcal{G}(i)\) is the set of pairs involving instance \(i\), and \(\theta\) is a temperature parameter. We apply UCL \cite{chen2020simple} to samples that are not part of confident pairs.

For robust representation learning, we also incorporate a Mixup technique \cite{zhang2018mixup}, blending pairs of samples as \( \boldsymbol{x}_i = \lambda \boldsymbol{x}_a + (1 - \lambda) \boldsymbol{x}_b \). Here, \(\lambda\) is a value between 0 and 1, following a Beta distribution. This Mixup approach is integrated into the contrastive loss, ensuring a balanced and effective learning process. Our system calculates the loss for each mixed example as a linear combination of the losses of the individual examples. The equation is written as:
\begin{equation}
\mathcal{L}_{\text{MIX}}(\boldsymbol{z}_i) = \lambda \mathcal{L}_a(\boldsymbol{z}_i) + (1 - \lambda) \mathcal{L}_b(\boldsymbol{z}_i),
\end{equation}
where the formulation of \( \mathcal{L}_a \) and \( \mathcal{L}_b \) is identical as the general contrastive loss \(\mathcal{L}\) described earlier. It's important to note that in our approach, the value of \( \lambda \) determines the dominant label for positive/negative pair selection for each mixed instance.

Furthermore, we use confident examples for adopting a classification learning objective to stabilize model convergence and improve representation quality. The classification loss \( \mathcal{L}_{\text{CLS}} \) is applied to these examples, contributing to the overall learning process.

\begin{equation}
\mathcal{L}_{\text{CLS}} = \sum_{(\boldsymbol{x}_i, \tilde{y}_i) \in \mathcal{T}} \text{loss}(\boldsymbol{\hat{p}}(\boldsymbol{x}_i), \tilde{y}_i),
\end{equation}

\noindent where "loss" denotes the loss function that measures the discrepancy between the predicted probabilities \( \boldsymbol{\hat{p}}(\boldsymbol{x}_i) \) and the confident labels \( \tilde{y}_i \), and \( \mathcal{T} \) constitutes the set of confident examples.

Our model also includes a learning objective that uses classifier predictions to learn similarity labels directly \cite{DBLP:journals/corr/abs-2006-07831}. That's how we treat the multi-viewed mini batch data, where we calculate a similarity loss \( \mathcal{L}_{\text{SIM}} \) to refine our model's performance further.

\begin{equation}
\mathcal{L}_{\text{SIM}} = \sum_{i \in I} \sum_{j \in A(i)} \text{loss}((\boldsymbol{\hat{p}}(\boldsymbol{x}_i)\boldsymbol{\hat{p}}(\boldsymbol{x}_j)), I[P_{ij} \in \mathcal{G}]),
\end{equation}

\noindent where \(I\) is an indicator function that determines if pairs \( (i, j) \) belong to the set of confident pairs \( \mathcal{G} \), and \(A(i)\) includes all indices except \(i\), enabling learning from diverse instance comparisons.

The culmination of our approach is encapsulated in the total objective loss, which combines these elements:
\begin{equation}
\mathcal{L}_{\text{ALL}} = \mathcal{L}_{\text{MIX}} + \lambda_c \mathcal{L}_{\text{CLS}} + \lambda_s \mathcal{L}_{\text{SIM}},
\end{equation}
where \( \lambda_c \) and \( \lambda_s \) represent  weights of the losses. We use \( \lambda_c = 1 \) and \( \lambda_s = 0.01 \) in all our experiments. This balanced framework fosters a positive feedback loop, in which finding confident pairs and enhancing representations mutually reinforce each other. 

\begin{comment}
\begin{figure}[t]
	\centering
	\includegraphics[width=\linewidth]{acmart-1.80-wisec2024_2/heatmap_gestures_drai.pdf}
	\caption{Single-frame heatmaps of six activities captured at a distance of 0.6m and an angle of 0°. }
	\label{fig:heatmap_activities_drai}
\end{figure}
\end{comment}

\subsection{Enhancing Classification with Fine-Tuning Technique}

In the final stage, we retain the previously trained CNN-LSTM-based deep encoder \(f\) and add a new classifier head for building a classifier network \(f_0\). This stage involves fine-tuning the complete model based on the confident activity samples by utilizing a customized version of Cross-entropy loss as a robust loss function \cite{DBLP:journals/corr/abs-2012-04462}.  Finally, the classifier  \(f_0\) outputs the activity recognition result.

%Overall, Our methodology employs a CNN-LSTM architecture tailored for robust activity classification using an mmWave-based activity dataset. This Selective  Supervised Contrastive Learning approach presents an innovative and effective solution for interpreting complex, time-dependent activity data from mmWave radar and solving the malicious label issue.

\section{Evaluation}
\begin{comment}
\begin{figure}[t]
	\centering
	\includegraphics[width=\linewidth]{acmart-1.80-wisec2024_2/heatmap_gestures_drai.pdf}
	\caption{Single-frame heatmaps of six activities captured at a distance of 0.6m and an angle of 0°. }
	\label{fig:heatmap_activities_drai}
\end{figure}
\end{comment}
\begin{comment}
\begin{figure}[htbp]
	\centering

		\centering
		\includegraphics[width=\linewidth]{acmart-1.80-wisec2024_2/testfig_HGR_CRNN_SL.pdf}
		\caption{Loss vs. Epoch and Accuracy vs. Epoch for SL.}
		\label{fig:test_hgr_crnn_sl}
\end{figure}

\begin{figure}[htbp]
		\centering
		\includegraphics[width=\linewidth]{acmart-1.80-wisec2024_2/LOSS_VS_EPOCH_ACCURACY_SCL.pdf}
		\caption{Loss vs. Epoch and Accuracy vs. Epoch for SCL.}
		\label{fig:loss_epoch_scl}
\end{figure}
\end{comment}

We first evaluate our prototype system with CNN+LSTM-based supervised learning (SL) and supervised contrastive learning (SCL). Then, we discuss the performance of the attacks on the prototype system. Lastly, we examine the effectiveness of the proposed defenses, which selectively employ samples for training of supervised contrastive learning (Sel-CL).

\subsection{Evaluation of the Prototype System}

We conduct experiments in various environments, including a meeting room, living room, bedroom, laboratory, and two distinct office rooms. These environments, characterized by their distinct sizes and furniture configurations, introduce diverse multipath effects, a crucial element that influences the performance of activity recognition systems. The data capture setup varied in terms of distance and angle across different types of locations, with configurations for locations 1 through 5 being: (0.6 meters, 0 degrees), (0.8 meters, 0 degrees), (1.0 meters, 0 degrees), (0.8 meters, -30 degrees), and (0.8 meters, 30 degrees), respectively. This variation adds an additional layer of complexity and realism to the activity classification. Training and testing of the system involve six activities: Push (PH), Pull (PL), Slide Left (LS), Slide Right (RS), Clockwise Turning (CT), and Anticlockwise Turning (AT).

\begin{figure}[htbp]

		\centering
		\includegraphics[width=0.8\linewidth]{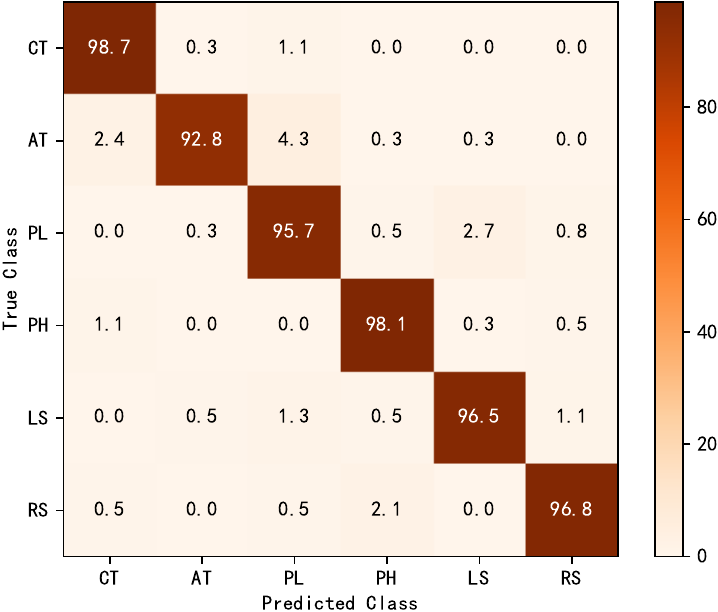}
		\caption{Test confusion matrix of SL-based HAR.}
		\label{fig:test_confusion_sl}

\end{figure}

\begin{figure}[htbp]
		\centering
		\includegraphics[width=0.8\linewidth]{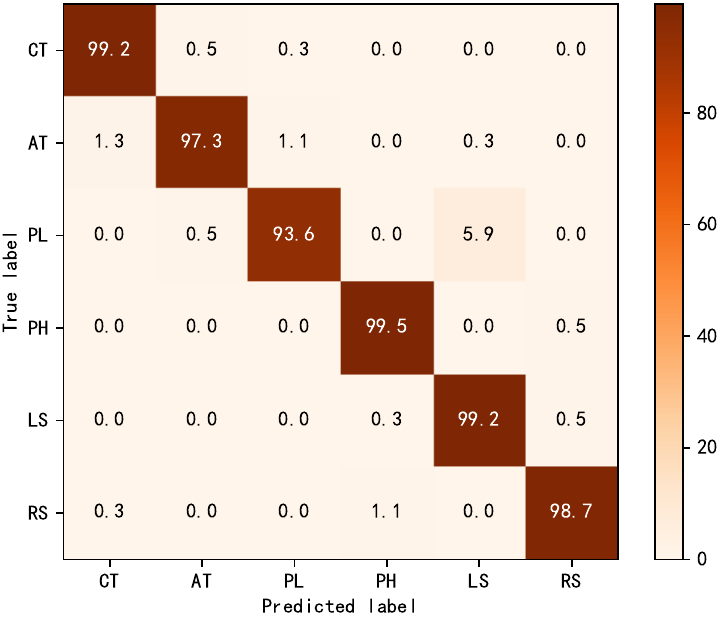}
		\caption{Test confusion matrix of SCL-based HAR.}
		\label{fig:test_confusion_scl}
\end{figure}

\begin{figure}[htbp]
    \centering
    \includegraphics[width=0.8\linewidth]{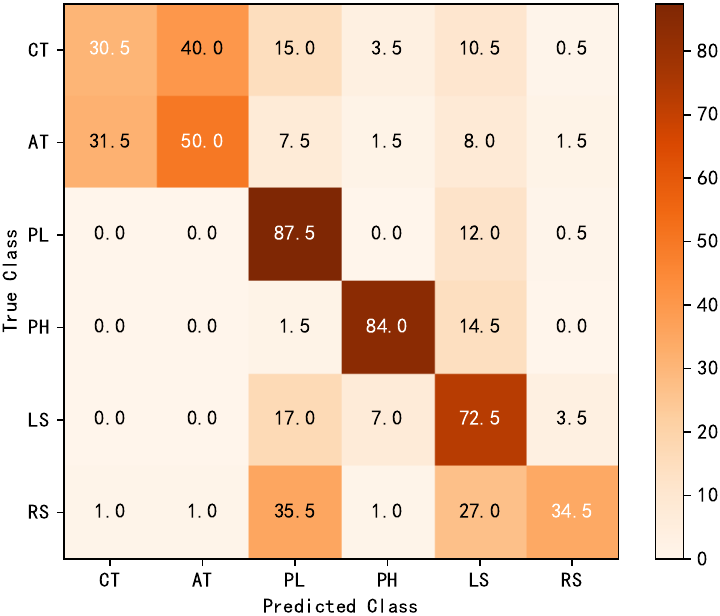}
    \caption{Test confusion matrix of SL-based HAR on extreme angle scenarios.}
    \label{fig:test_confusion_sl_la}
\end{figure}

\begin{figure}[htbp]
    \centering
    \includegraphics[width=0.8\linewidth]{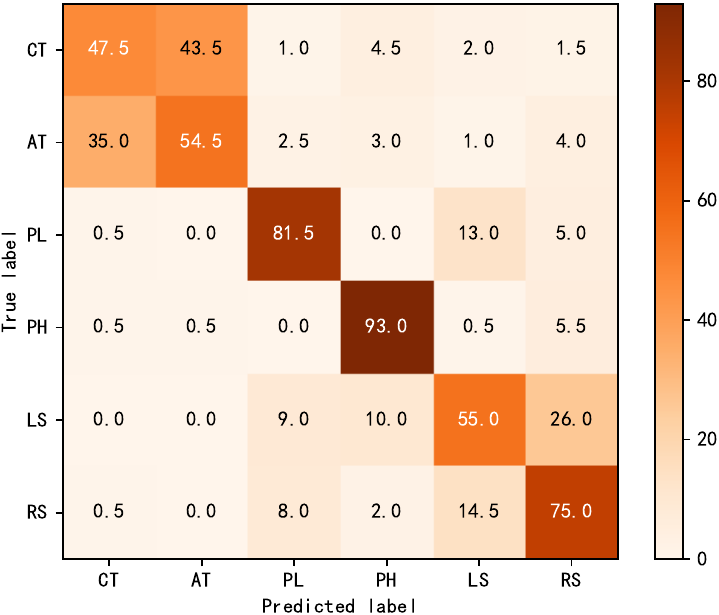}
    \caption{Test confusion matrix of SCL-based HAR on extreme angle scenarios.}
    \label{fig:test_confusion_scl_la}
\end{figure}

\begin{figure*}[t]
\centering
% SL row
\begin{minipage}{.32\textwidth}
  \centering
  \includegraphics[width=\linewidth]{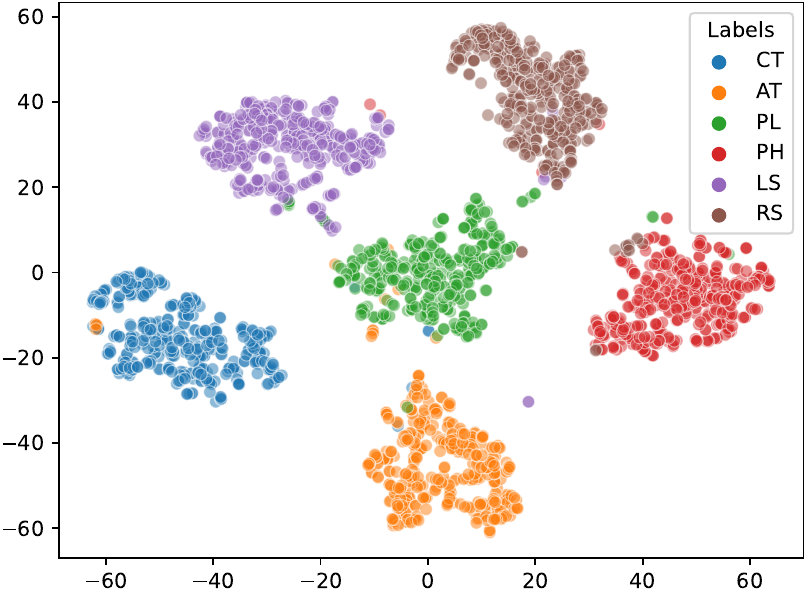}
  \subcaption{SL with clean data.}
  \label{fig:SLclean}
\end{minipage}%
\hfill
\begin{minipage}{.32\textwidth}
  \centering
  \includegraphics[width=\linewidth]{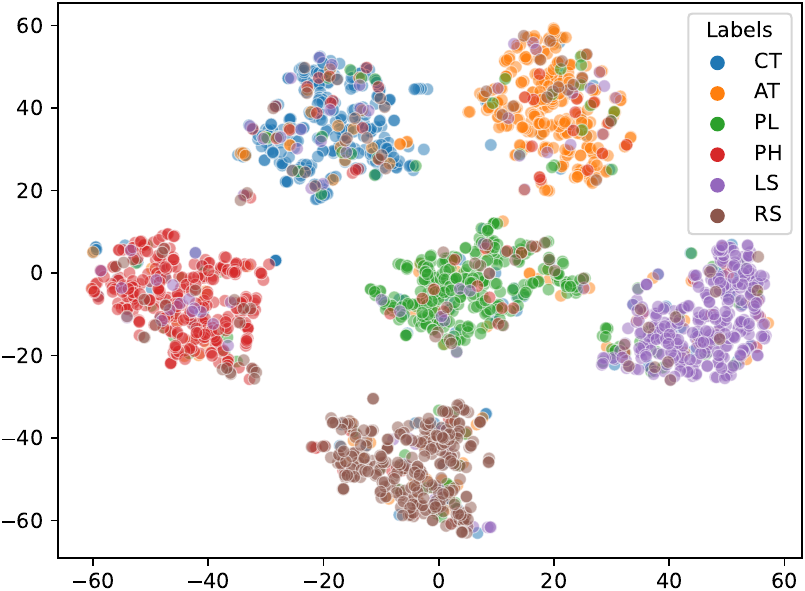}
  \subcaption{SL under 20\% random attack.}
  \label{fig:SL20}
\end{minipage}%
\hfill
\begin{minipage}{.32\textwidth}
  \centering
  \includegraphics[width=\linewidth]{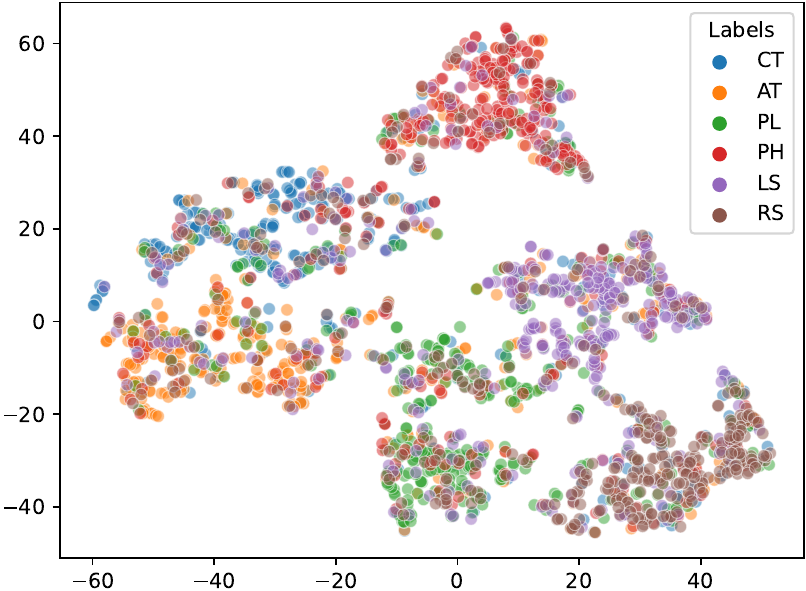}
  \subcaption{SL under 40\% random attack.}
  \label{fig:SL40}
\end{minipage}
\caption{Feature representations for SL-based HAR with clean training data and under random attacks.}
\label{fig:sl_rep}
\end{figure*}

\begin{figure*}[t]
\centering
% SCL row
\begin{minipage}{.32\textwidth}
  \centering
  \includegraphics[width=\linewidth]{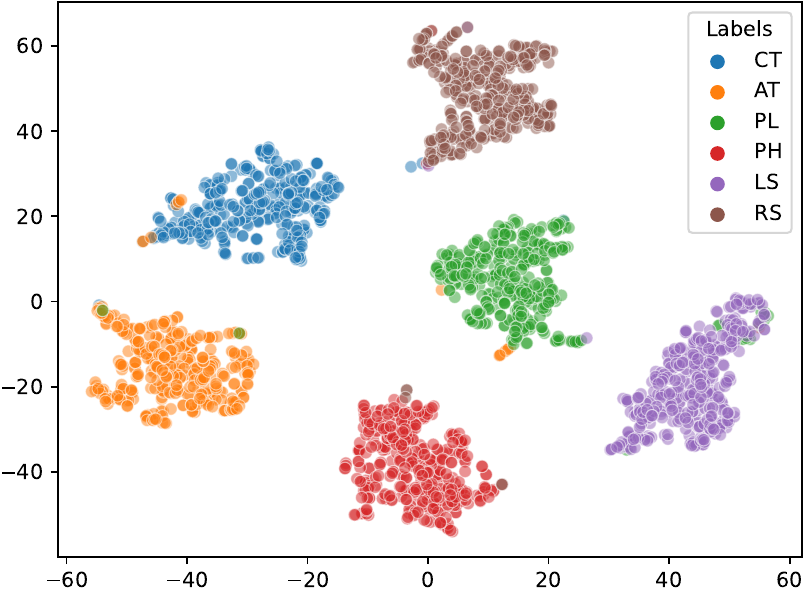}
  \subcaption{SCL with clean data.}
  \label{fig:SCLclean}
\end{minipage}%
\hfill
\begin{minipage}{.32\textwidth}
  \centering
  \includegraphics[width=\linewidth]{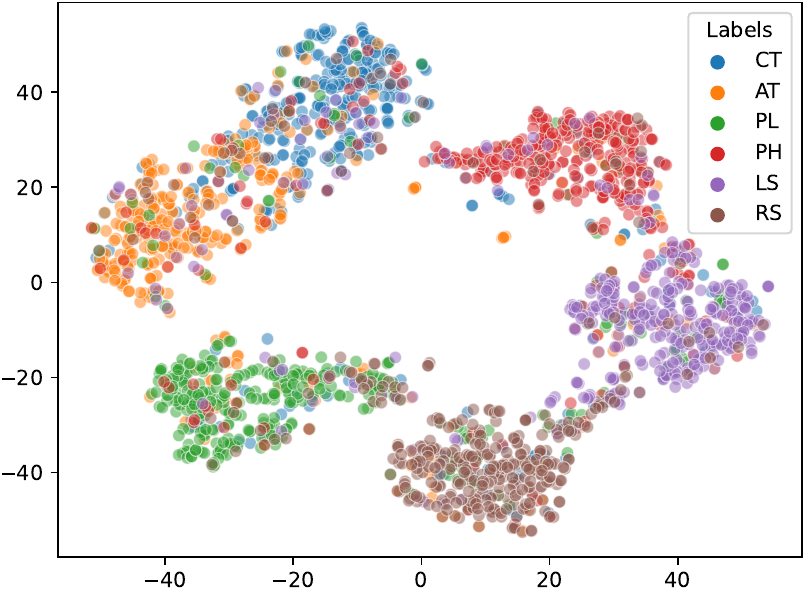}
  \subcaption{SCL under 20\% random attack.}
  \label{fig:SCL20}
\end{minipage}%
\hfill
\begin{minipage}{.32\textwidth}
  \centering
  \includegraphics[width=\linewidth]{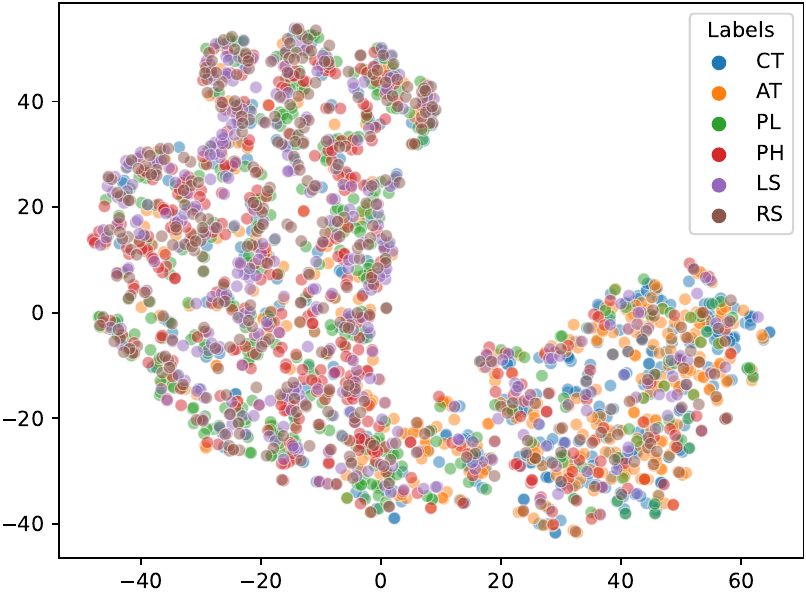}
  \subcaption{SCL under 40\% random attack.}
  \label{fig:SCL40}
\end{minipage}
\caption{Feature representations for SCL-based HAR with clean training data and under random attacks.}
\label{fig:scl_rep}
\end{figure*}

\begin{figure*}[t]
\centering
% SEL-CL row
\begin{minipage}{.32\textwidth}
  \centering
  \includegraphics[width=\linewidth]{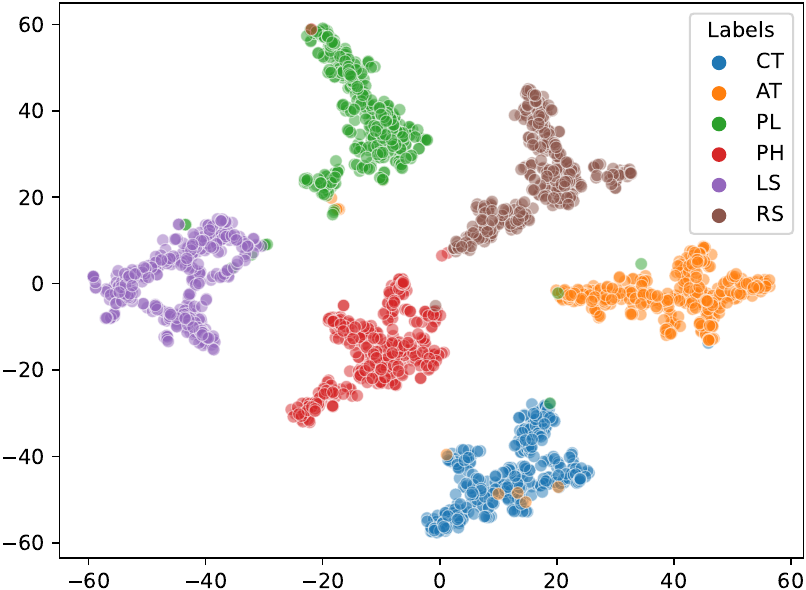}
  \subcaption{Sel-CL with clean data.}
  \label{fig:SELclean}
\end{minipage}%
\hfill
\begin{minipage}{.32\textwidth}
  \centering
  \includegraphics[width=\linewidth]{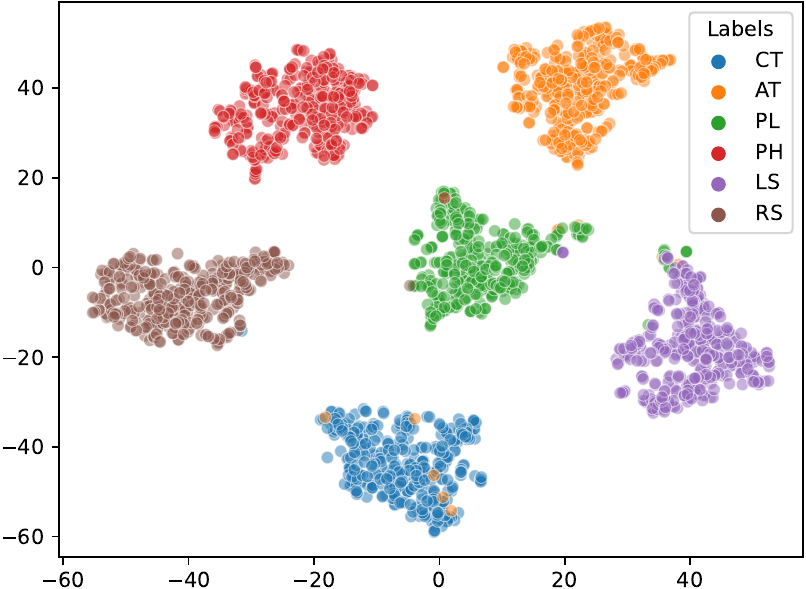}
  \subcaption{Sel-CL under 20\% random attack.}
  \label{fig:SEL20}
\end{minipage}%
\hfill
\begin{minipage}{.32\textwidth}
  \centering
  \includegraphics[width=\linewidth]{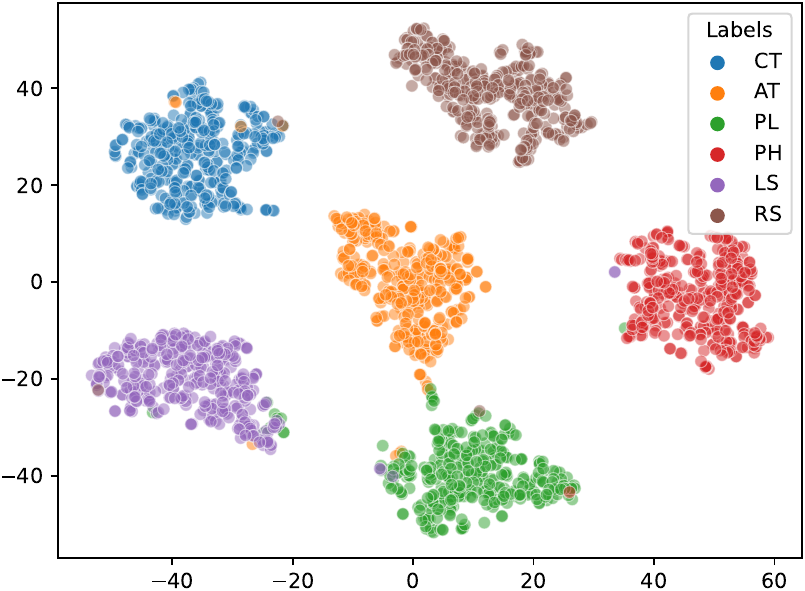}
  \subcaption{Sel-CL under 40\% random attack.}
  \label{fig:SEL40}
\end{minipage}
\caption{Feature representations for Sel-CL-based HAR with clean training data and under random attacks.}
\label{fig:selcl_rep}
\end{figure*}
%The dataset encompasses six distinct activity types, each represented by a 3D array. The arrays' first dimension varies slightly, corresponding to the temporal length of the activity, while the remaining two dimensions (32×32) are consistent across all activities. This variation reflects the different durations or complexities of each activity. The consistent 32×32 dimensions in the other two axes suggest a uniform resolution of the captured data across all activities. The first dimension of each array represents a temporal sequence, with each slice in this dimension being a snapshot at a particular time point which we can consider as a frame. The 32×32 dimensions represent spatial, or feature data captured for each time point or frame. 

%Figure~\ref{fig:heatmap_activities_drai} displays 2D heatmaps for 

%The six activities used in training and testing are Push (PH), Pull (PL), Slide Left (LS), Slide Right (RS), Clockwise Turning (CT), and Anticlockwise Turning (AT).  These heatmaps are extracted from the middle frame of each activity, captured at a distance of 0.6m and an orientation of 0°, within a controlled meeting room setting. This image offers a snapshot that reflects the typical patterns observed in each class. The unique patterns in each heatmap highlight the feasibility of mmWave-based HAR.

For our study, we divided the dataset into training, validation, and test sets based on the environment of data collection. The training set, comprising 6,300 samples, was derived from the bedroom, laboratory, and two office rooms. The meeting room data, with 2,100 samples, formed our validation set, while the living room data, consisting of 2,250 samples, was used as the test set.

\subsubsection{Performance in Normal Conditions}

We first evaluate system performance under normal conditions, with the user's angle approximately at 0°. Figure \ref{fig:SLclean} and Figure \ref{fig:SCLclean} below show the feature representations learned by our model using the SL and SCL approaches, respectively. These visualizations illustrate the clear distinction and separation of activity representations when using SCL, compared to the more entangled representations in SL.  Our SCL model achieves a remarkable 97.92\% testing accuracy, demonstrating SCL's effectiveness in activity classification. In contrast, a parallel evaluation with the SL-based approach yields a 96.43\% accuracy, 1.49\% lower than SCL. This differential not only underscores the superiority of SCL in our context but also aligns with existing literature \cite{khosla2021supervised}.
\begin{comment}

\begin{figure}[htbp]
	\centering
    \begin{minipage}{.5\textwidth}
		\centering
		\includegraphics[width=\linewidth]{acmart-1.80-wisec2024_2/features_sl.pdf}
		\caption{Features Representation using SL}
		\label{fig:features_sl}
    \end{minipage}
    \hfill
	\begin{minipage}{.5\textwidth}
		\centering
		\includegraphics[width=\linewidth]{acmart-1.80-wisec2024_2/features_scl.pdf}
		\caption{Features Representation using SCL}
		\label{fig:features_scl}
	\end{minipage}%
\end{figure}

Figure \ref{fig:test_hgr_crnn_sl}  and Figure \ref{fig:loss_epoch_scl} offer a comparative analysis of the loss versus epoch and accuracy versus epoch in the training processes for SL and SCL, respectively. It is evident from these figures that SCL not only converges faster but also achieves a higher accuracy more consistently across epochs.  
\end{comment}
Figure \ref{fig:test_confusion_sl} and Figure \ref{fig:test_confusion_scl} show the test confusion matrices for SL and SCL models. The SCL model outperforms the SL model in almost every class.

%The incorporation of Supervised Contrastive Learning (SCL) in mmWave-based activity recognition marks a substantial leap toward enhanced robustness and precision. In sectors ranging from home automation, and industry to healthcare, the accuracy of activity recognition is of paramount importance\cite{8303728, 10.1145/3432235}. The improvement of almost 1.5\% in accuracy, while seemingly modest, has profound real-world consequences. In critical domains such as automated patient monitoring and industrial automation, where precision is crucial, this incremental enhancement in accuracy by our proposed SCL based mmWave activity recognition system will significantly bolster reliability and reduce the potential for errors. Such improvements are crucial in elevating operational efficiency and improving safety standards. Our results indicate that SCL is not just an incremental upgrade but a pivotal advancement in the field, dramatically enhancing the reliability and effectiveness of mmWave-based activity recognition systems.

\subsubsection{Performance in Extreme Conditions}

We also evaluate the system's performance in extreme conditions, specifically when the user deviates from the radar's center. The HAR system, trained with data collected under normal conditions, is tested with 10 users in extreme conditions at angles of ±45° and ±60°. That is, the testing data is entirely unseen during the training phase. The SL-based system achieves an accuracy of 59.83\% in these challenging scenarios. Again, the SCL-based system outperforms the SL-based system, achieving an accuracy of 67.75\%, which is 7.92\% higher than the SL-based model.  Confusion matrices illustrated in Figures \ref{fig:test_confusion_sl_la}  and \ref{fig:test_confusion_scl_la} show that the SCL model outperforms SL model in extreme angle scenarios for most of the classes.

%Figures \ref{fig:features_la_sl} and \ref{fig:features_la_scl} show feature representations in extreme angles using SCL and SL. SCL demonstrates superiority over SL in the representation of features within extreme angle scenarios.

\begin{comment}

\begin{figure}[htbp]
	\centering
    \begin{minipage}{.5\textwidth}
        \centering
        \includegraphics[width=\linewidth]{acmart-1.80-wisec2024_2/features_large_anglel_sl.pdf}
        \caption{Features Representation using SL for extreme angle scenario}
        \label{fig:features_la_sl}
	\end{minipage}
    \hfill
	\begin{minipage}{.5\textwidth}
		\centering
		\includegraphics[width=\linewidth]{acmart-1.80-wisec2024_2/features_scl_large_angle.pdf}
		\caption{Features Representation using SCL for extreme angle scenario}
		\label{fig:features_la_scl}
	\end{minipage}%
\end{figure}
\end{comment}

\begin{figure*}[htbp]
\centering
\begin{subfigure}[b]{0.32\textwidth}
    \includegraphics[width=\textwidth]{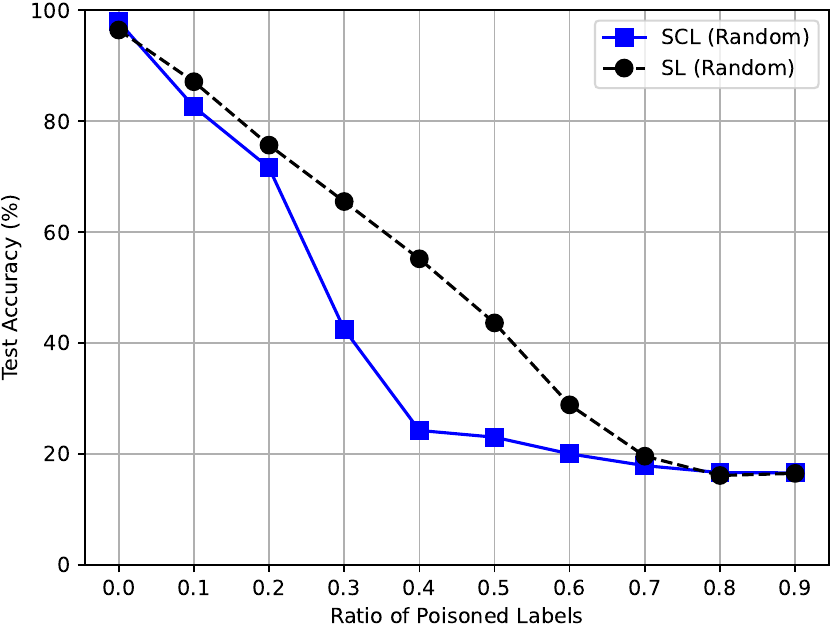}
    \caption{Random attack.}
    \label{fig:random-attack}
\end{subfigure}
\hfill
\begin{subfigure}[b]{0.32\textwidth}
    \includegraphics[width=\textwidth]{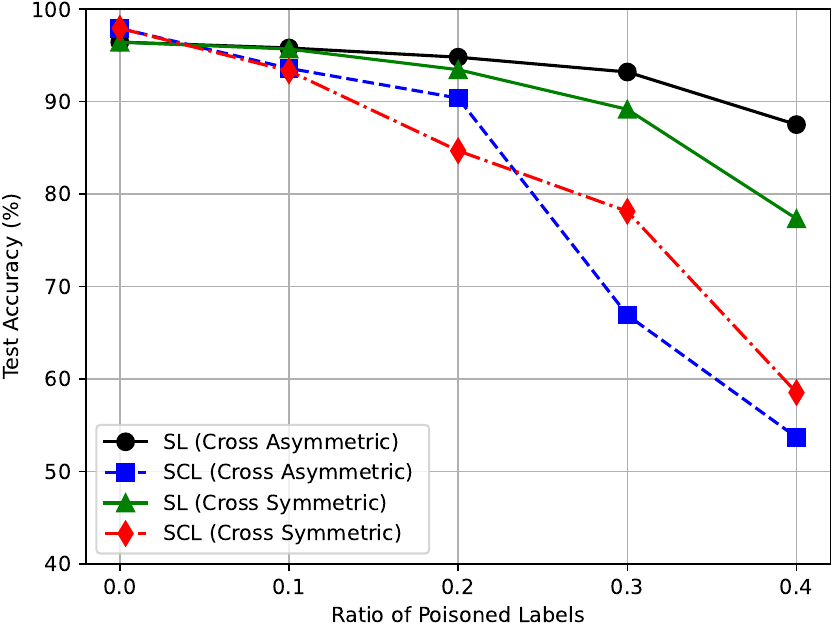}
    \caption{Cross trajectory attack.}
    \label{fig:cross-trajectory-attack}
\end{subfigure}
\hfill
\begin{subfigure}[b]{0.32\textwidth}
    \includegraphics[width=\textwidth]{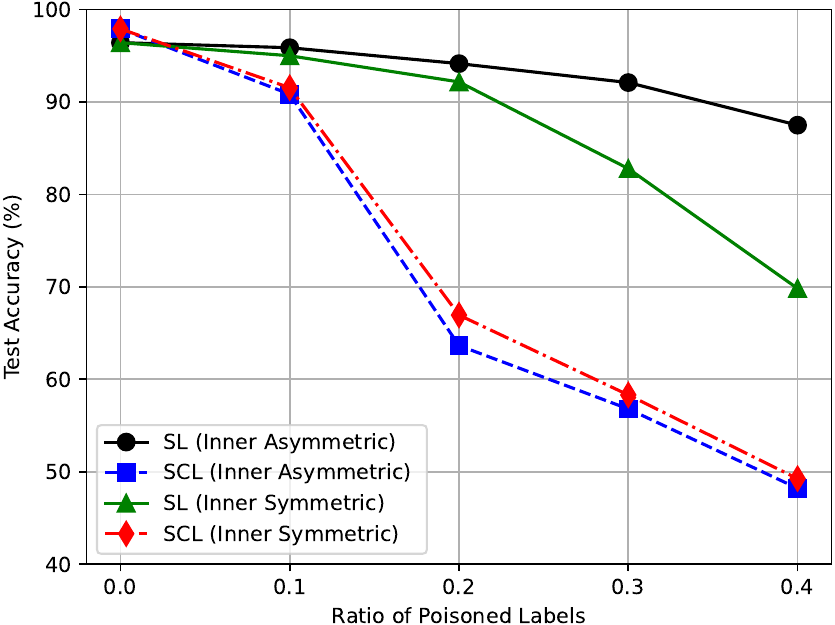}
    \caption{Inner trajectory attack.}
    \label{fig:inner-trajectory-attack}
\end{subfigure}
\caption{Impact of three label poisoning attacks on SL and SCL based HAR systems.}
\label{fig:attacks}
\end{figure*}

\subsection{Evaluation of Attacks}

\begin{figure}[htbp]
    \centering
    \includegraphics[width=0.9\linewidth]{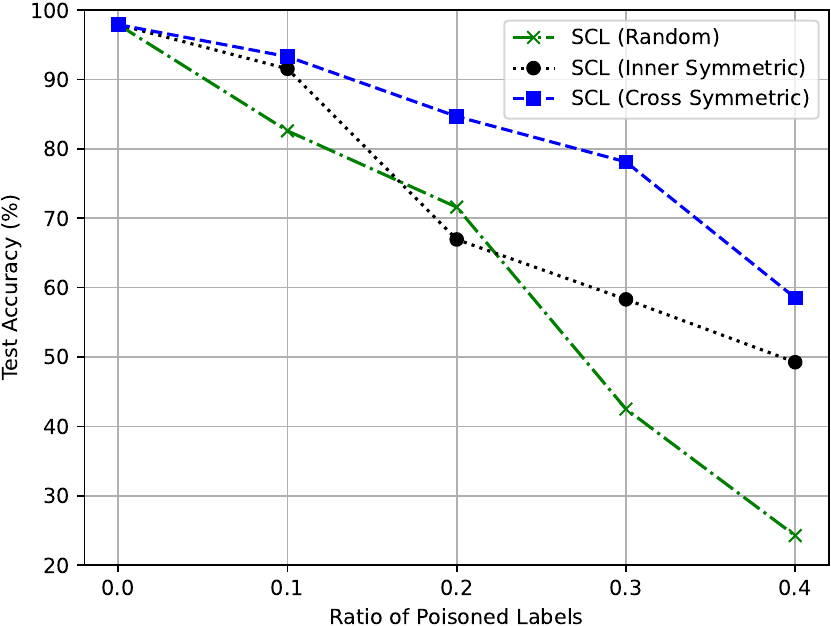}
    \caption{Comparing three attacks on SCL-based HAR.}
    \label{fig:threeattacksSCL}
\end{figure}

Figure \ref{fig:sl_rep} and \ref{fig:scl_rep} present feature representations of SL and SCL based HAR under random attacks. As the ratio of poisoned labels increases, instances of the activities become increasingly intermixed with one another.  We can observe that SCL is more vulnerable to poisoned labels than SL. With 40\% poisoned labels, differentiating instances of activities becomes nearly impossible. 

Figure \ref{fig:random-attack} illustrates the classification accuracy when a random attack is adopted. The accuracy of the HAR based on both SL and SCL drops significantly with the increasing ratio of poisoned labels, aligning with the previous feature representation figures. The accuracy of both models drops to around 70\% with malicious label ratio of 20\%. We can observe that the performance of SCL-based HAR is better than SL-Base HAR with clean training data but also collapses faster in the presence of malicious labels. This result demonstrates that malicious labels significantly affect the generation of positive and negative pairs in SCL. It also implies that SCL-based systems require enhanced protection against label flipping attacks.

Figure \ref{fig:cross-trajectory-attack} shows classification accuracy in the presence of cross trajectory attacks. The accuracy of the system declines as the number of poisoned labels increases, although not as drastically as with random attacks. Again, the SCL-based HAR system exhibits increased vulnerability to these malicious labels. In addition, symmetric flipping attacks exert a more pronounced effect on the system's performance, owing to the involvement of a greater number of classes in the attack. Both models become nearly unusable with a malicious label ratio of above 30\% in symmetric-based attacks.

Figure \ref{fig:inner-trajectory-attack} shows that the system accuracy under inner trajectory attacks drops faster compared to cross trajectory attack especially for the SCL models. The findings indicate that activities sharing similar trajectories significantly influence the generation of accurate positive and negative examples in contrastive learning compared to those with dissimilar trajectories. 

\begin{figure*}[htbp]
\centering
\begin{subfigure}[b]{0.32\textwidth}
    \includegraphics[width=\textwidth]{0.4_tsne_random_attack_scl.pdf}
    \caption{Random Attack.}
    \label{fig:randomrepresentation}
\end{subfigure}
\hfill
\begin{subfigure}[b]{0.32\textwidth}
    \includegraphics[width=\textwidth]{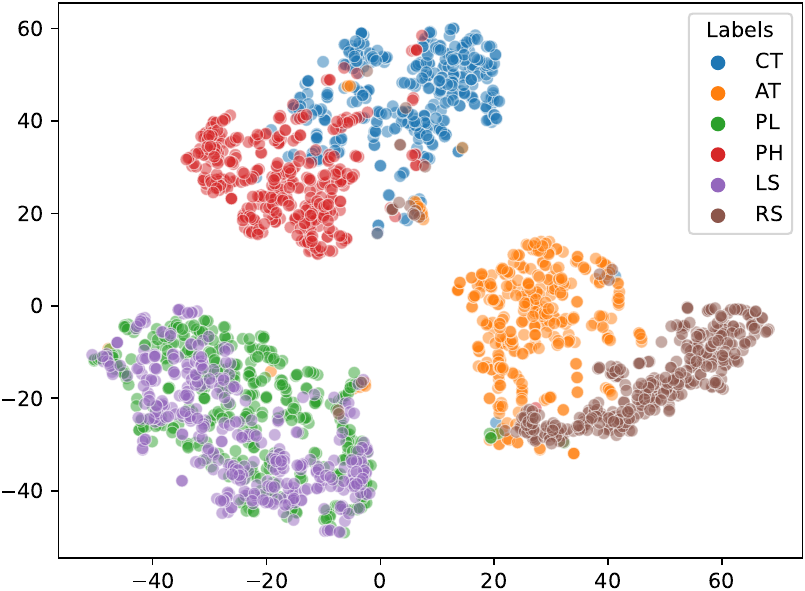}
    \caption{Cross Trajectory Attack.}
    \label{fig:crossrepresentation}
\end{subfigure}
\hfill
\begin{subfigure}[b]{0.32\textwidth}
    \includegraphics[width=\textwidth]{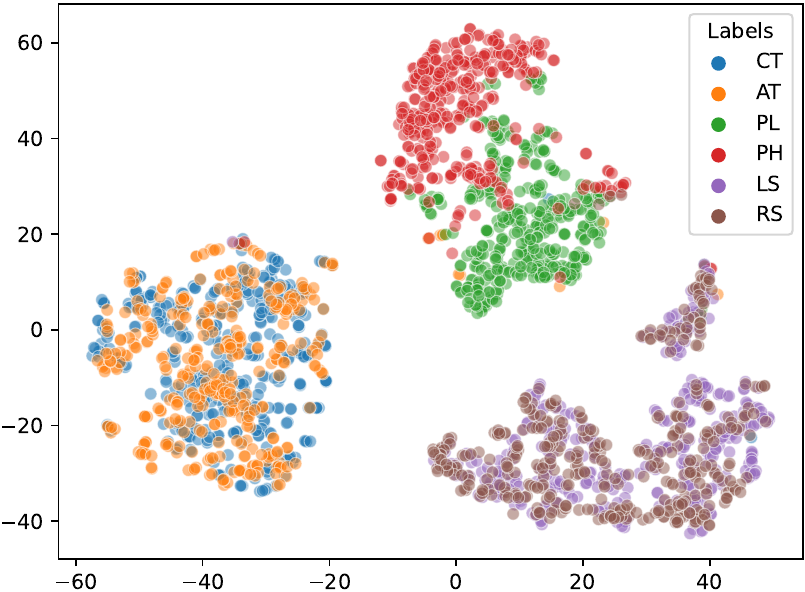}
    \caption{Inner Trajectory Attack.}
    \label{fig:innerrepresentation}
\end{subfigure}
\caption{Feature representations of SCL-based HAR under three different attacks with poisoned label ratio of 40\%.}
\label{fig:attacks_rep}
\end{figure*}

Figure \ref{fig:threeattacksSCL} compares the impact of the three attacks on the SCL-based HAR system. It is observed that random attacks are more effective than both inner and cross trajectory attacks, with the inner trajectory attack proving to be more potent than the cross trajectory attack. Figure \ref{fig:attacks_rep} demonstrates the consistent impacts of three attacks with a 40\% poisoned label ratio on feature representations. Both cross and inner trajectory attacks are conducted in symmetric forms. Under random attacks, instances of different activities are completely intermingled. The feature representation for the cross trajectory attack is marginally better than that for the inner trajectory attack.

Based on the observations, it is clear that label flipping attacks can significantly impact the accuracy of both models, highlighting the need for effective defenses. While the SL model may not perform as well as the SCL model in the absence of attackers, it shows greater resilience to various types of attacks. Furthermore, random attacks that poison the training data with both similar and dissimilar trajectories are the most potent. Inner trajectory attacks, which involve adding labels for similar trajectories into the training data, offer a strategic method for attackers aiming to conceal malicious label tampering while still intending to make a notable impact.

\subsection{Evaluation of Defenses}

\begin{figure}[ht]
	\centering
	\includegraphics[width=0.9\linewidth]{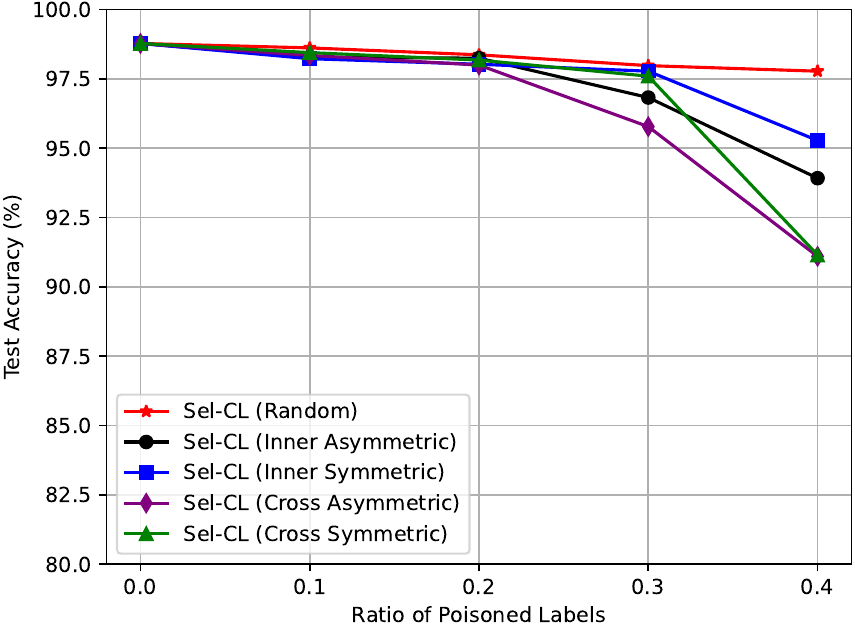}
	\caption{Performance of Sel-CL in defending against various attacks.}
	\label{fig:defense_Sel_CL}
\end{figure}

\begin{figure}[ht]
	\centering
	\includegraphics[width=0.9\linewidth]{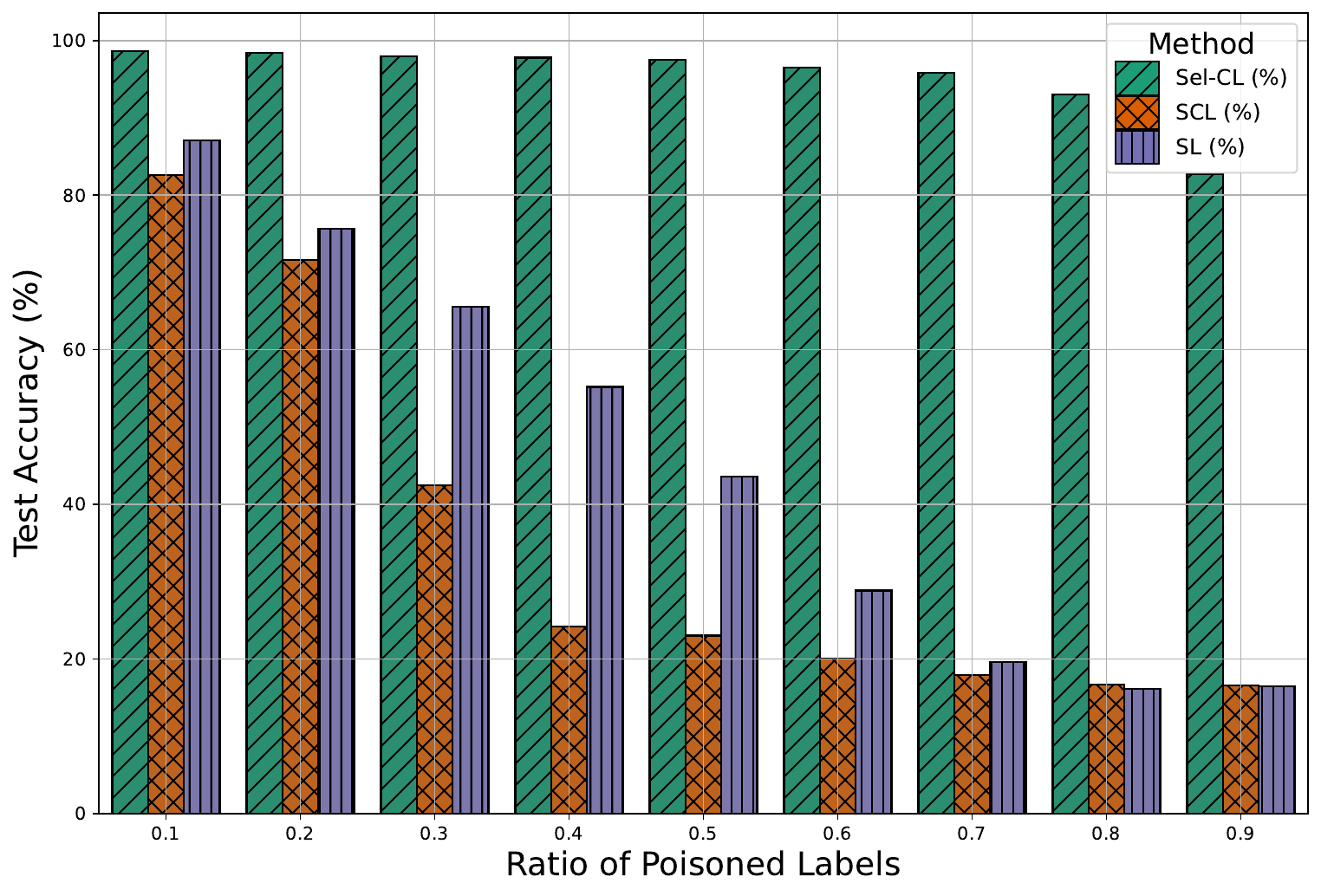}
	\caption{Accuracy of three techniques under random attack. }
	\label{fig:bar_plot_across_methods}
 \vspace{-.1in}
\end{figure}

\begin{figure}[ht]
	\centering
	\includegraphics[width=0.9\linewidth]{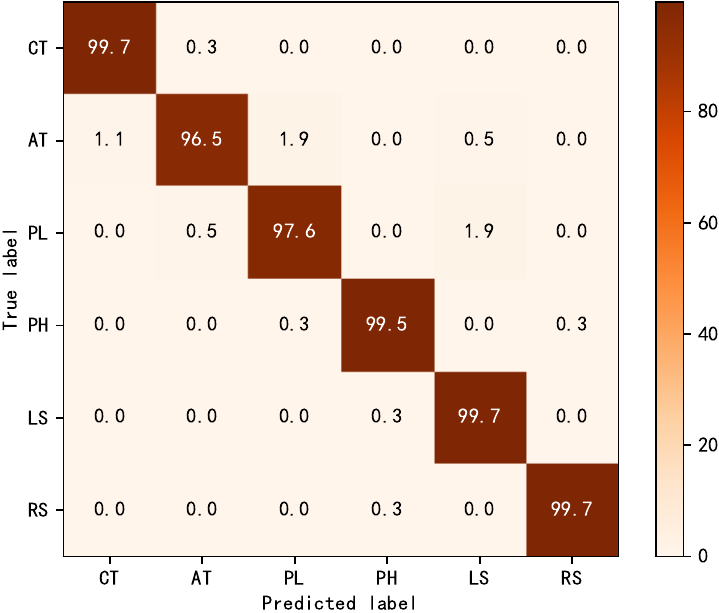}
	\caption{Test Confusion Matrix of Sel-CL-based HAR. }
	\label{fig:0.0_selcl_random_attack_cm}
\vspace{-.1in}
\end{figure}

Figure \ref{fig:selcl_rep} illustrates the feature representations with clean data and under random attacks when we employ the Sel-CL technique as the defense. It is observed that instances of different activities remain distinctly separable, even with a poisoned label ratio of 40\%. Figure \ref{fig:defense_Sel_CL} demonstrates the impressive capability of Sel-CL in defending against all previously mentioned attacks. The Sel-CL-based HAR system maintains nearly the same accuracy level with up to 20\% malicious labels for all types of attacks. Remarkably, even when the malicious label ratio increases to 40\%, Sel-CL still maintains an accuracy of above 90\% for all attacks. Surprisingly, the random attack, which significantly impacts SL and SCL models, has minimal effect on Sel-CL. The cross and inner trajectory attacks, which employ fixed label manipulation methods, can induce larger errors in the Sel-CL model. From these findings, it is evident that random label flipping has a lesser impact on generating confident examples compared to fixed label manipulation.

Figure \ref{fig:bar_plot_across_methods} compares the performance of Sel-CL, SCL, and SL models with malicious label ratios up to 90\% under random attacks. When the ratio reaches 80\%, the Sel-CL-based system still achieves an accuracy of above 90\%, whereas the performance of SCL and SL-based systems deteriorates to a level comparable to random guessing.

Figure \ref{fig:0.0_selcl_random_attack_cm} displays the test confusion matrix for the HAR system utilizing Sel-CL with clean training data. The Sel-CL-based system demonstrates excellent performance across all activities, achieving an accuracy of 98.78\%, which exceeds the performance of SCL (97.92\%) and SL (96.43\%) under similar conditions. These results lead to the conclusion that the defense mechanisms do not adversely affect system performance when no attacks are present. On the contrary, the defenses actually enhance accuracy by filtering out low-quality samples during the training data selection process.

\section{Related Work}

Wireless human activity recognition (HAR) technology has attracted significant attention due to its non-intrusive nature. Past research has developed wireless HAR systems based on various techniques such as Received Signal Strength Indicator (RSSI) \cite{7765094}, Doppler Profiles \cite{PuWho13} and Channel State Information (CSI) \cite{9490685, YAN2023119042,10.1145/2536853.2536873}. Recently, as a breakthrough in wireless communication, mmWave technology has also been used to build high-resolution HAR systems \cite{10230151, article, s23218901, 9858146, 10001175}. However, past research focuses on improving performance of the wireless HAR systems, neglecting the security issues in the systems.

%Nowadays, Gesture recognition using millimeter-wave (mmWave) technology has gained attention due to its superior capabilities compared to Channel State Information (CSI) and Received Signal Strength Indicator (RSSI) based systems\cite{9490685, 7218525, 7765094, 10.1145/2536853.2536873}. The performance of CSI and RSSI-based systems is often limited due to environmental noise and lower resolution, which results in less accurate recognition of gestures\cite{YAN2023119042, 10.1145/2536853.2536873, 10.1145/3131672.3131692, 10.1145/3210240.3210335}. mmWave technology solves these limitations by utilizing Frequency Modulated Continuous Wave (FMCW) radar which is non-intrusive, works well in the dark, and has a high resolution for fine-grained gesture recognition, making it a promising choice for diverse applications in the presence of obstacles and non-line of sight scenarios\cite{9860976, s23218901, 10230151}.  mmWave Radar is also preferable than the camera and wearable sensors based gesture recognition due to its privacy preserving capabilities, minimal use of energy, and noncontactness\cite{article, s23218901, 9858146}. 

In addition to new wireless sensing techniques, machine learning mechanisms, such as supervised learning \cite{9002720, 1048206}, semi-supervised learning \cite{YAN2023119042, 10.5555/3172077.3172113} and contrastive learning \cite{s23063332, Singh_2021_CVPR}, have been a critical part for activity recognition. 
Most existing studies focus on applying unsupervised contrastive learning to reduce the labeling cost in wireless HAR systems \cite{SongRfu22}.
A notable advancement by Khosla \textit{et al.} \cite{khosla2021supervised} shows that supervised contrastive learning provides higher accuracy than traditional supervised learning algorithms \cite{Cunningham2008} on classification tasks. Contrastive learning focuses on representation learning by pulling similar class samples together and pushing the dissimilar class samples apart. This paper is the first to apply supervised contrastive learning technique for mmWave-based HAR and protect it from label flipping attacks.

%Basically, for gesture classification deep learning model is used, whereas some literature utilizes Supervised Learning (SL)\cite{9002720, 1048206}, others used semi-supervised learning\cite{YAN2023119042, 10.5555/3172077.3172113} and contrastive learning\cite{s23063332, Singh_2021_CVPR}. A notable advancement by khosla et al.\cite{khosla2021supervised} shows that Supervised Contrastive Learning (SCL) provides higher accuracy than traditional Cross entropy-based Supervised Learning algorithms\cite{Cunningham2008} on classification tasks. Contrastive learning, in its supervised form, has emerged as a promising approach. In contrastive learning, the model focuses on representation learning by pulling similar class samples together and pushing the dissimilar class samples apart and contrastive learning can be used to build effective neural network models\cite{chen2020simple, Liu_2021, Singh_2021_CVPR}. SCL is based on the contrastive learning idea but utilizes the label information to effectively generate positive and negative pairs for learning representations. Hence, the classifier task becomes easier to classify\cite{khosla2021supervised}.

Deep learning-based systems for activity recognition are susceptible to label flipping poisoning attacks, which aim to compromise system performance through the insertion of malicious labels. \cite{HanAdv12, XiaoSup15} designed optimized label flipping attacks that target support vector machines (SVM). However, the mechanisms are tailored for SVM and cannot be used directly for attacking neural networks. Some papers \cite{ShahiLab22, ShahiAss23} also present label flipping attacks for wearable HAR systems, yet they do not take into account trajectory similarity during the attack process. Furthermore, they assume that there exists a trusted training dataset to train a preliminary model for the defense. In addition to the label flipping attacks, researchers in the machine learning field also studied noisy label issues, which mainly address unintentional mislabeling \cite{Xiao_2015_CVPR, 9729424, li2022selectivesupervised, algan2020label, 10.1007/978-981-15-0694-9_38, DBLP:journals/corr/ZhangBHRV16}. Current studies on noisy labels do not consider attacking strategies and primarily focus on general machine learning algorithms, neglecting the distinct characteristics in mmWave-based HAR. 

This paper is the first one to identify trajectory-based label flipping attacks on mmWave-based HAR systems in the context of supervised contrastive learning and to propose corresponding defenses without depending on any trusted dataset.

\section{Conclusion}

In this paper, we presented the first systematic study on the vulnerabilities of mmWave-based Human Activity Recognition (HAR) systems that rely on supervised contrastive learning. We identified three novel label flipping attacks by considering trajectory similarity and evaluated their impact on our prototype system. Additionally, we developed defenses by prioritizing confident examples during the training process. Extensive experiments demonstrated that our defenses are highly effective against these attacks. The attack and defense strategies introduced in this paper can be readily extended to other wireless HAR systems.

\section{Acknowledgements}
This work was supported in part by the U.S. National Science Foundation under grants CNS-2422863, CNS-2325563, and CNS-2055751. Research was also sponsored in part by the Army Research Laboratory and was accomplished under Cooperative Agreement Number W911NF-23-2-0225. The views and conclusions contained in this document are those of the authors and should not be interpreted as representing the official policies, either expressed or implied, of the Army Research Laboratory or the U.S. Government. The U.S. Government is authorized to reproduce and distribute reprints for Government purposes notwithstanding any copyright notation herein.

\bibliographystyle{ACM-Reference-Format}
% \bibliography{mmrobust_references}
\balance
\bibliography{mmrobust_references,nets22}

	%, nets22, wins, winspub}

\end{document}